\documentclass[journal]{IEEEtran}
\usepackage{microtype}
\usepackage{graphicx}
\usepackage{subfigure}
\usepackage{booktabs} % for professional tables
\usepackage{amsfonts}
\usepackage{tikz}
\usepackage{changepage}
\usepackage{gensymb}
\usepackage{amsmath}
\usepackage{algorithm}
\usepackage{algorithmic}
\usepackage{pgf}
\usepackage[utf8x]{inputenc}
\usetikzlibrary{arrows.meta, bending, backgrounds, scopes}

\usepackage{hyperref}

\ifCLASSINFOpdf
\else
\fi
\begin{document}
% Copyright notice for posting on arXiv and personal website
\onecolumn
© 2020 IEEE.  Personal use of this material is permitted.  Permission from IEEE must be obtained for all other uses, in any current or future media, including reprinting/republishing this material for advertising or promotional purposes, creating new collective works, for resale or redistribution to servers or lists, or reuse of any copyrighted component of this work in other works.
\twocolumn
\title{A hybrid learning method for system identification and optimal control}

\author{Baptiste~Schubnel,
		Rafael~E.~Carrillo,~\IEEEmembership{Member,~IEEE,}
        Pierre-Jean~Alet,~\IEEEmembership{Senior~Member,~IEEE,}
        and~Andreas~Hutter %~\IEEEmembership{Life~Fellow,~IEEE}% <-this % stops a space
\thanks{The authors are with CSEM, Neuch\^{a}tel, Switzerland (email: baptiste.schubnel@csem.ch; rafael.carrillo@csem.ch; pierre-jean.alet@csem.ch; andreas.hutter@csem.ch).}% <-this % stops a space
}


\markboth{IEEE Transactions on Neural Networks and Learning Systems, Accepted}%
{Shell \MakeLowercase{\textit{et al.}}: Bare Demo of IEEEtran.cls for IEEE Journals}



\maketitle

\begin{abstract}
  
We present a three-step method to perform system identification and optimal control of non-linear systems. Our approach is mainly data driven and does not require active excitation of the system to perform system identification. In particular, it is designed for systems for which only historical data under closed-loop control are available and where historical control commands exhibit low variability.    In a first step, simple simulation models of the system are built and run under various conditions. In a second step, a neural network architecture is extensively trained on the simulation outputs to learn the system physics, and retrained with historical data from the real system with stopping rules. These constraints avoid overfitting that arise by fitting closed-loop controlled systems. By doing so, we obtain one (or many) system model(s), represented by this architecture, and whose behaviour can be chosen to match more or less the real system.  Finally, state-of-the-art reinforcement learning with a variant of domain randomization and distributed learning is used for optimal control of the system.  We first illustrate the model identification strategy with a simple example, the pendulum with external torque. We then apply our method to model and optimize the control of a large building facility located in Switzerland. Simulation results demonstrate that this approach generates stable functional controllers which outperform on comfort and energy benchmark rule-based controllers.
\end{abstract}


% \fi
\IEEEpeerreviewmaketitle

%\IEEEPARstart{T}{his} demo
\section{Introduction}
\label{Introduction}
Control of complex systems involves both system identification and controller design, which are challenging and time consuming tasks if the system exhibits complex non-linear dynamics. Model Predictive Control (MPC)  finds the control law to a system as the solution of an optimization problem \cite{camacho2007}. MPC has a long deployment history in the chemical control process  industry, for which it has been originally developed \cite{di2012industry}. Most of the  industrial modern  model predictive controllers deployed  are based on linear systems of equations, for which the system identification method  is well established (see e.g., \cite{van2012subspace}) and the optimization methods under constraints are theoretically well grounded (see e.g., \cite{borrelli2017predictive}). Widespread and large scale deployment in other fields is hindered by well-known limitations of MPC: model identification requires engineering and fine tuning, that can be hard to scale, especially if the systems to control exhibit a large diversity of configurations \cite{mayne2014}. For fields where the processes exhibit non-linearities, both system identification and control can be cumbersome and require important engineering work and fine tuning-steps (see e.g., \cite{schoukens2019nonlinear}). Furthermore, system identification can be really compromised if system excitations are hindered for security or operative reasons. In that case, most of the system identification methods will lead to over fitted models and will not generalize well to new input commands. The aforementioned situation is encountered in the field of energy management systems for the building industry, on which one of our test cases is built. Modern systems like heat pump or HVAC (Heating, Ventilation and Air Conditioning) exhibit non-linear dynamics \cite{jones2007air}, and, for large building facilities, building operators and technicians are often reluctant to carry out system excitations that are needed for traditional system identification, as they may impact occupants comfort. 

We propose here a three-step method that is tailor-made for system identification and optimal control of non-linear systems, where excitations for system identification cannot be carried out and only historical data under closed loop control are available.    In a first step, an approximate simulation model of the system is built. This simulation model can be rough and based on simple and generic system blocks, reproducing  the essential physics. For instance, it can rely on simplified/linearized equations of the true dynamic system with coarse estimation of parameters. 

This simplified simulation model is used in the second step to perform system identification. In order to learn reduced models of the system,  a neural network architecture is first trained extensively on data generated with the simplified simulation and then shortly retrained on real historical data via transfer learning techniques. The obtained reduced models are able to capture the physics of the systems relatively well. Different training strategies lead to different models, that we accumulate to build an ensemble of reduced models. We base our reduced models on recurrent neural networks (RNN), due to their ability to fit both linear and non-linear functions, but also because they are relatively simple to retrain and well-suited for transfer learning tasks. 

The third and last step is the design of the optimal controller(s).   The reduced models obtained in the second step of the method could be directly used for optimal control in a MPC-framework. However, if neural networks models are used, one has to resort to non-linear optimization techniques. In various cases, especially if the number of control variables is high, the optimization programs may require high computational resources to be deployed in a real time optimization setup.  Moreover, they also rely on heuristics and have no theoretical convergence guarantees in most practical problems. In this article, we take another path and make use of the ensemble of reduced models produced via the second step of our method to learn a controller that is trained ``offline'' to control in an optimal way all elements in the ensemble in parallel. We train this controller using a recently developed on-policy reinforcement learning algorithm, PPO (Proximal Prolicy Optimization, \cite{schulman2017}), that has shown good results in real robotic control tasks \cite{ppo_18}. To enforce robustness, we train the controller on many variants of the reduced model, a procedure called domain randomization in the reinforcement learning literature; see e.g., \cite{tobin2017domain} and \cite{peng2018sim}. Reinforcement learning has in general two main drawbacks: it is very sample inefficient and there are, to our knowledge,  no theoretical convergence guarantees for most of the recent methods using non-linear function approximators, such as deep neural networks. However, some recent works show a positive trends in that direction, see e.g., \cite{dai2017sbeed} and \cite{liu2019neural}. One of its strengths is that it is very cheap in computational resources once deployed, with the load of computational resources being shifted to the training phase. Moreover, it can be trained so as not  to rely on environmental predictions (but on its collected experiences and past values) for its control strategy.

The main contributions of our approach are:
\begin{itemize}
\item A new neural network architecture capable of reproducing the dynamics of both linear and non-linear systems,
\item A training process based on transfer learning techniques to identify the aforementioned architecture with real systems dynamics and to minimize overfit on closed loop historical data,
\item A distributed method to learn robust controllers for optimal control of the identified system.
\end{itemize}
Our approach has the main advantage to be purely offline, in the sense that system identification and controller design do not require any direct intervention on the real system before deployment. In particular, controller validation tests can also be performed in an offline fashion, prior to deployment.

The organization of the article is as follows. In section \ref{soa}, we review the state of the art for system identification, model predictive control and reinforcement learning. In section \ref{sec2}, we give an in-depth description of our identification and  control learning method. In section \ref{pendulum_case}, we test the system identification method on a simple example, the pendulum with external torque. We evaluate the full approach, system identification and optimal control, by applying our method to control a large office building in Switzerland. Simulation results are presented in section \ref{building}. Finally, we conclude in section \ref{conclusion} with closing thoughts and future directions.

\section{State of the art}
\label{soa}
Linear system identification is a well-established field with standard references like \cite{ljung2001system} and \cite{van2012subspace}. Popular implementations for linear state space models  exist in many programing languages, like the N4SID algorithm in Matlab identification toolbox. These algorithms are based on the singular value decomposition of the product of the extended observability matrix with the state. Linear Kalman filters are used for state estimation and tracking. However, most systems of interest are nonlinear or only linear to a first approximation \cite{schoukens2016}. Non-Linear system identification methods relies on two main approaches: a physical approach, where simulated models are constructed based on the non-linear physical equations underpinning the system dynamics; and a ``black box'' approach, where non-linear models like NARX or Wiener-Hammerstein model are used (see e.g., \cite{schoukens2016,schoukens2019nonlinear} and references therein for further details on the subject). There has been a growing interest in recent years in using neural networks for system identification, and many architectures have been presented in the literature \cite{nechyba1994,ogunmolu2016,gonzales2018}. A reduced model architecture similar to ours (see section \ref{sec2}) is used in \cite{baumeister2018deep}.   In \cite{champion2019data}, the authors propose to use autoencoders and sparse regression techniques to reconstruct sparse dynamic equations from real system data.   
 
MPC is among the most versatile and used model-based control approaches. It involves an online optimization of the control strategy over a prediction receding horizon \cite{camacho2007}. The optimization problem allows the inclusion of tailored objective functions and constraints on system behaviour and variables. MPC relies on dynamic models of the system to be controlled. These models are used to predict the response of the system to a control signal inside the optimization algorithm. The model also takes into account external signals and past plant states and outputs. The optimization is defined over the interval $[t,t+H]$,  where $t$ is the current time and $H$ is the prediction (optimization) horizon. Only the first (discrete time) step of the solution is implemented, then the plant state is sampled again, and a new optimization is repeated in a receding horizon fashion. Most MPC approaches consider simple, linear state space models to model the system dynamics \cite{mayne2014}, though data-driven approaches that combine the representational power of deep neural networks with the flexible optimization framework of MPC have also been proposed in the literature \cite{lenz2015,baumeister2018deep,drgona2018,verma2018,bieker2019}.  

In the field of building control, MPC provides a powerful framework to exploit energy storage capabilities and optimization of renewable energy sources generation (see \cite{serale2018} and references therein). Adoption of MPC was limited until a decade ago due to its high computational demand \cite{mocanu2019}, though with the development of new processors and computing resources, MPC is increasingly applied in various types of buildings and energy systems \cite{lindelof2015,serale2018}. One of the main challenges in MPC is the trade-off between model accuracy and model simplicity such that the optimization problem is computationally tractable while achieving meaningful control strategies \cite{mayne2014}. Nowadays, modern machine learning methods, such as reinforcement learning, can overcome the on-line computation limitation by  automatically computing (optimizing) the control strategy and shifting the computational load to an off-line training phase.
 
Reinforcement learning (RL) is an old field (see e.g., \cite{sutton2018reinforcement}) that has attracted a growing interests in recent years because of successful approaches using neural networks as function approximators: Deep Q-learning for video game playing \cite{mnih2015human},  combination of policy gradients, self-play and tree search for mastering the game of go \cite{silver2016mastering}, and on-policy methods for real-time multiplayer games like Starcraft \cite{vinyals2019alphastar}. However, most of these great results were accomplished on games or simulated environments, for which interactions between the learning agent(s) and the environment can be carried out millions of times: RL with neural networks as function approximators is sample inefficient. To overcome sample inefficiency, several approaches are currently intensively investigated: among them, one can cite imitation learning (see e.g., \cite{schaal1999imitation} and \cite{osa2018algorithmic}) for robotics or autonomous vehicles, for which human demonstrations can be provided;  model-based RL, where explicit or latent environment models are used in the training phase, resulting in better sample efficiency than model-free methods; see e.g., \cite{vezzani2019learning}, \cite{janner2019trust} for recent developments in the field. Our approach in this paper is close to domain randomization and to the approach followed in  \cite{peng2018sim} and \cite{ppo_18}. In these papers, however, complete simulated environments were provided, a clear obstacle for large-scale industrial applications. In our approach, we alleviate the simulation work by using reduced models that are trained on basic simulations and then retrained on historical data; this method is presented in Section \ref{sec2}.

\section{Method description}
\label{sec2}
\begin{figure}[ht]
\begin{center}
\centerline{\includegraphics[scale=0.85]{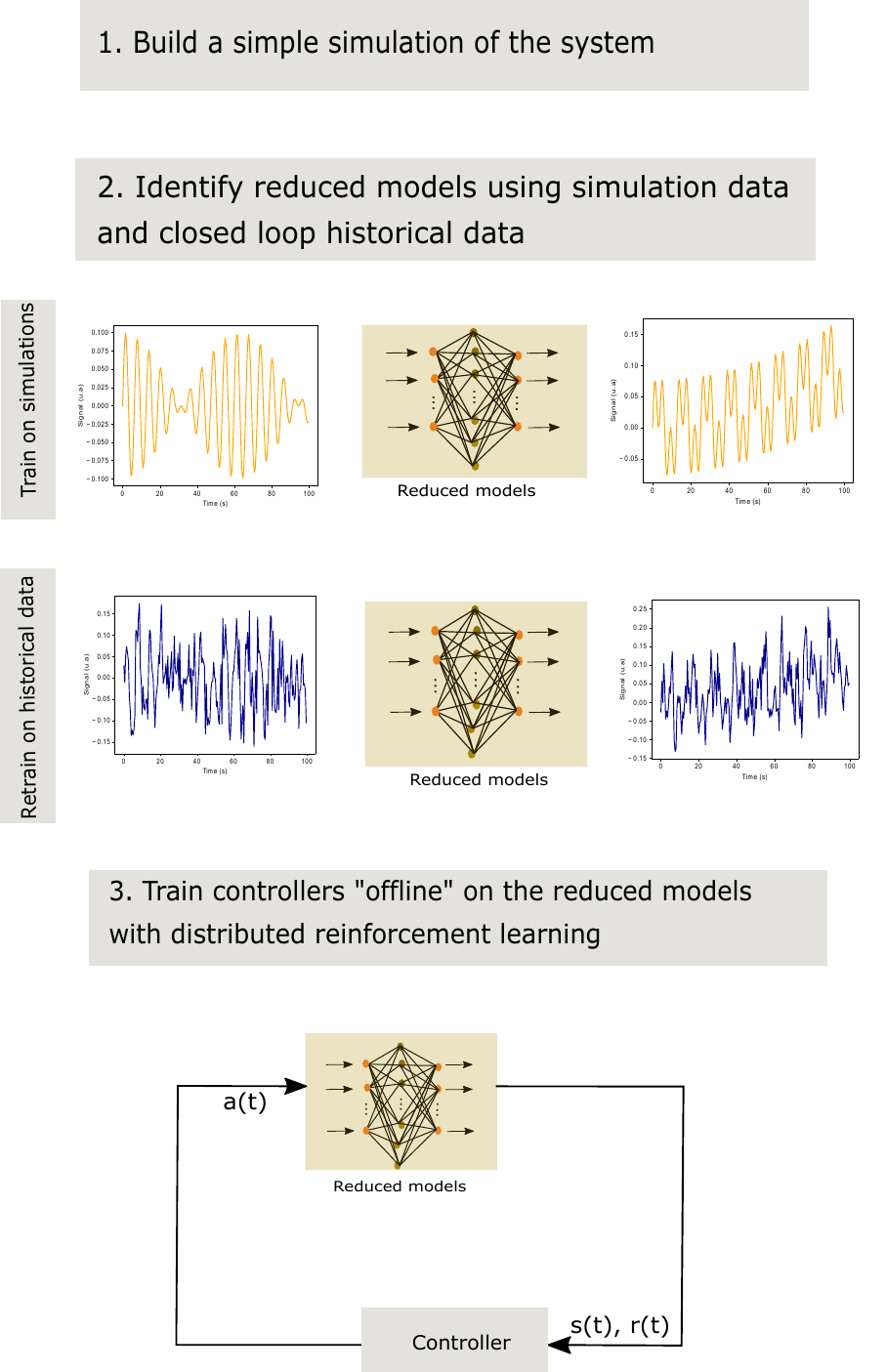}}
\caption{  Illustration of the three-step method. Step 1: build a simple simulation model and use it to generate data that serves as a prior source of information of the system dynamics. Step 2: reduced models (neural networks) with wide approximation capabilities are first trained on these data and then adapted and fine tuned on the available but restricted real historical data. Step 3: reduced models from step 2 are used in a third step to train controllers.}
\label{fig_process}
\end{center}
\end{figure}

We describe below our three-step method for system identification and control.  The three steps are illustrated in Figure \ref{fig_process}.  

\subsection{Step 1:  Semi-physical Simulation}
\label{simple_sim}
The first step consists in building a simple simulation model of the dynamic system.   This model is used to generate data that incorporate some prior but inexact information of the system dynamics and that will be used to pre-train the reduced models in the second step. The model has to be based on a partial knowledge of the physics underlying the system dynamics or the effective dynamics of the system variables of interest. Our method is  hence designed for systems, that can be locally approximated by a set of partial differential equations. This approach is dubbed semi-physical modeling in the literature (see e.g., \cite{schoukens2019nonlinear}, p. 18), in the sense that building the semi-physical simulation does not require a precise and detailed mathematical model of the system, but rather a reduced set of equations that captures its main dynamical features.    For example, in the pendulum test case, we construct a toy simulation based on the linearized version of true dynamics. For the large facility building case, we rely on resistance capacitance models for buildings (see e.g., \cite{bueno2012resistance}, hereafter RC-models), that represent temperatures evolution in rooms and building elements with first order linear differential equations, in analogy with elementary electrical laws. More details about the tools and assumptions used for the building simulation are given in Section \ref{building}. 

\subsection{Step 2: System Identification}
\label{nn}

%\noindent \textbf{Reduced model architecture.}
\subsubsection*{Reduced model architecture} 
We propose to use a general encoder-decoder architecture where both the encoder and the decoder are based on Long Short-Term Memory (LSTM) cells \cite{lstm_art}. This network architecture, inspired by natural language processing architectures \cite{NIPS2014_5346}, is adapted to the special needs of dynamic systems: the encoder is an LSTM-network that is used to initialize the hidden state of the system, whereas the decoder is another LSTM network followed by a multilayer perceptron (MLP) that  is used to predict the model outputs step by step.   The use of LSTM networks to represent dynamic systems is justified by their properties i.e., unlike plain RNNs, they have been designed to handle both long and short term dependencies via a gating mechanism (see e.g., \cite{lstm_art}, where the role of the cell state as information keeper is discussed). Moreover, the use of non-linear activation functions makes them suitable to capture non-linearities. Other recent neural network architectures sharing these properties (e.g., stacked LSTMs, gated recurrent units \cite{cho2014properties} or transformer architectures \cite{vaswani2017attention}) could be used as well.   The encoder plays a similar role to a Kalman filter for hidden state estimation: given a sequence of observed commands and outputs, it creates a representation of the system under control at time $t_0$. This representation may encode hidden degrees of freedom that play a role in the observed dynamic variables, e.g., walls, air flows, etc. for buildings.    This intuition is supported by the link between linear state space model representation and RC models of buildings , see e.g., \cite{fateh2019state}  . The decoder takes the initial state representation and updates the observations based on input commands.

Let $n,l \in \mathbb{N}$ be the number of steps used to initialize the model and the number of steps in the prediction horizon, respectively.  Denote by $x(t)$ the tuple of command inputs, exterior parameter values ( e.g., outside temperature and irradiance for a building) and system observables (e.g., rooms temperature for a building) at time step $t$, with total dimension $d = d_{I} + d_{E} + d_{O} $ at every step, and by  $x^{\sharp}(t)$ the tuple of command inputs and exterior parameter values  at step $t$, with total summed dimension $d^{\sharp} = d - d_{O} $. The observations are denoted by $o(t) \in \mathbb{R}^ {d_{O}}$. A schematic view of the  neural architecture is given on Figure \ref{nnfig}.

\def\layersep{1.5cm}
\begin{figure}[ht]
\begin{adjustwidth*}{}{0em} 
    \begin{tikzpicture}[shorten >=1pt,->,draw=black!50, 
    node distance = \layersep,
every pin edge/.style = {<-,shorten <=1pt},
        neuron/.style = {circle,fill=black!25,minimum size=17pt,inner sep=0pt},
  input neuron/.style = {neuron, fill=green!50},
 output neuron/.style = {neuron, fill=red!50},
 hidden neuron/.style = {neuron, fill=blue!50},
         annot/.style = {text width=4em, text centered},
                         ]

 \node[input neuron, pin=left: $x(t_{0}-n)$] (I-1) at (-2,-1) {};
 \node[input neuron, pin=left: $x(t_{0}-n+1)$] (I-2) at (-2,-2) {};
 \node[input neuron, pin=left: $x(t_{0}-1)$] (I-4) at (-2,-4) {};
 \node[input neuron, pin=left:  $ \ \vdots \ $ ] (I-3) at (-2,-3) {};

% Draw the decoder layer nodes
   \path[yshift=0.5cm]
    node[hidden neuron, pin=left: $x^{\sharp}(t_{0})$] (H-5) at (-0.5,-5-0.5) {};
   \path[yshift=0.5cm]
    node[hidden neuron, pin=left: $x^{\sharp}(t_{0}+1)$] (H-6) at (-0.5,-6-0.5) {};
   \path[yshift=0.5cm]
    node[hidden neuron, pin=left: $ \ \vdots \  $] (H-7) at (-0.5,-7-0.5) {};
   \path[yshift=0.5cm]
    node[hidden neuron, pin=left: $x^{\sharp}(t_{0}+l-1)$] (H-8) at (-0.5,-8-0.5) {};

\node[output neuron,pin={[pin edge={->}]right: $\hat{o}(t_{0})$}, right of=H-5] (O-5) {};
\node[output neuron,pin={[pin edge={->}]right: $\hat{o}(t_{0}+1)$}, right of=H-6] (O-6) {};
\node[output neuron,pin={[pin edge={->}]right: $  \vdots \  $ }, right of=H-7] (O-7) {};
\node[output neuron,pin={[pin edge={->}]right: $\hat{o}(t_{0}+l-1)$}, right of=H-8] (O-8) {};

%Connect every node in the input layer with every node in the hidden layer.
 \path (I-1) edge (I-2);
 \path (I-2) edge (I-3);
 \path (I-3) edge (I-4);

 \path (H-5) edge (H-6);
 \path (H-6) edge (H-7);
 \path (H-7) edge (H-8);

% Connect every node in the decoder layer with Fc layer
\foreach \source in {5,...,8}
       \path (H-\source) edge (O-\source);

% Connect every node in the hidden layer with the output layer
%\foreach \source in {1,...,5}
 %   \path (H-\source) edge (O);

% Annotate the layers
 \node[draw] at (-2, -0.1)   (enc) {Encoder};
 \node[draw] at (-0.5, -4.1)   (dec) {Decoder};
 \node[draw] at (1.3, -4.1)   (fc) {MLP};

% Annotate some arrows to indicate cell states 
 \node (c) at (0.25, -4.8)  {\scriptsize $h(t_0)$};
 \node (c) at (0.25, -5.8)  {\scriptsize $h(t_0+1)$};
 \node (c) at (0.25, -7.8)  {\scriptsize $h(t_{0}+l-1)$};

% curve from encoder to decoder
\scoped[on background layer]
\draw[line width=1mm, gray!50, shorten >=1mm, shorten <=1mm, -{Latex[flex]}]  (I-4) --  (H-5);
    \end{tikzpicture}

\end{adjustwidth*}
\caption{Neural Network architecture.  The vertical descending arrows correspond to the transfer of the tuples $(c(t),h(t))$ from time $t$  to time $t+1$ at every timestep $t$. The large descending arrow between the encoder and the decoder corresponds to the initialization of the decoder cell state and output with the encoder values $(c(t_0 -1), h(t_0 -1))$.   }
\label{nnfig}
\end{figure}
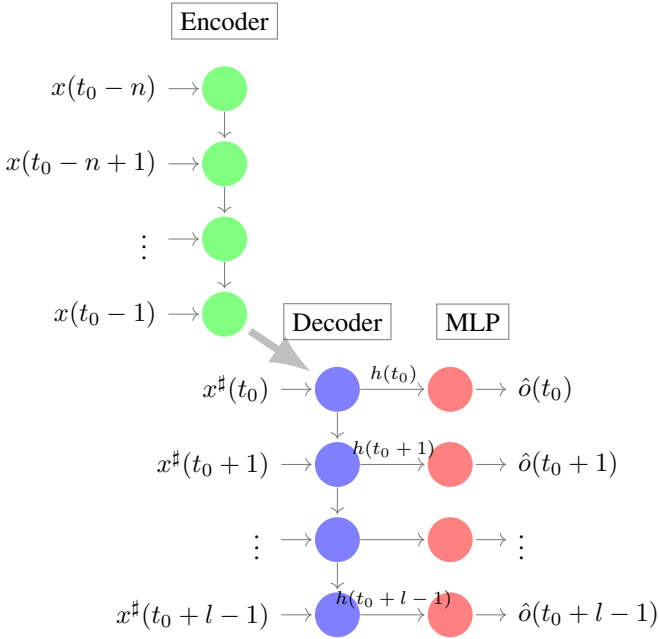

We denote by $c_{\tiny enc}(t) \in \mathbb{R}^{p}$ and $h_{\tiny enc}(t) \in \mathbb{ R}^{p}$  the encoder cell state and output (for gated recurrent units, only $h_{\tiny enc}(t)$ is used). The LSTM-encoder is used to generate the representation of the initial system state at time $t_0$, given the past $n$ values. It can be viewed as an iterative application of the map  $f_{\text{\tiny enc}} : \mathbb{R}^{d+2p}  \rightarrow \mathbb{R}^{2p}$ such that the tuple
\begin{equation}
\label{rec}
(c_{\tiny enc}(t),h_{\tiny enc}(t)) = f_{\tiny enc}(x(t),c_{\tiny enc}(t-1),h_{\tiny enc}(t-1))
\end{equation}
is a representation of the system state at time $t_0-1$ when $t=t_0-1$ and Eq. \ref{rec} has been repeated $n$ times from $t_0-n$ to $t_0-1$.  For $t' \geq t_0$, the decoder LSTM is a map $ f_{\tiny dec} : \mathbb{R}^{d^{\sharp}+2p} \rightarrow \mathbb{R}^{2p}$ with 
\begin{equation}
\label{rec2}
(c_{\tiny dec}(t),h_{\tiny dec}(t)) = f_{\tiny dec}(x^{\sharp}(t),c_{\tiny dec}(t-1),h_{\tiny dec}(t-1))
\end{equation}
that only depends on the commands and external parameters (but not previous system observables $o$), and where $c_{\tiny dec}(t_0-1) = c_{\tiny enc}(t_0-1)$ and  $h_{\tiny dec}(t_0-1) = h_{\tiny enc}(t_0-1)$. The output  $h_{\tiny dec}(t)$   is  fed to a MLP network which estimates the output $\hat{o}(t)$ at time $t$:
\begin{equation}
\hat{o}(t) = f_{MLP}(  h_{\tiny dec}(t) ).
\end{equation}

%\noindent \textbf{Training on simulation.}
\subsubsection*{Reduced model training}
 
As depicted on Figure \ref{fig_process}, reduced model training is carried out in two stages. 

In the first stage, the encoder, decoder and fully connected layers are trained on data generated via the simple system simulation described in Section \ref{simple_sim}. This training stage is carried out on a large dataset from the simulation. Its main purpose is to initialize the network weights with representations that already capture some essentials of the system dynamics. Training is also carried out with varying encoding and decoding sequence lengths to yield models with good prediction results for a wide range of forecast horizons. Small variations in the encoding length are kept, whereas large variations in the decoding length $l$ are used. The networks are trained in a supervised fashion with the square-loss function $\vert \hat{o}(t)-o(t) \vert^2$, where $o(t)$ is the array of system observations at time $t$. Stochastic gradient descent with the Adam optimizer as well as norm gradient clipping \cite{zhang2019gradient} are used for training. 

In the second stage, the encoder-decoder architecture is retrained on historical data from the real system. This retraining stage has to be performed with care since most data obtained from real systems are under closed loop control. Good examples are systems with PID controllers, that are controlled around a fixed setpoint or in a very limited range of setpoints. In that case, any model relying solely on historical data will exhibit poor generalization results to new control input patterns (see e.g., \cite{peters2013}, where causal methods did not yield good results in finding causal predictors for building rooms dynamics). Our strategy to overcome this issue is to rely on stopping rules, that we design to get reduced models with good generalization properties. The stopping rules are built on a set of environmental and command values, $\mathcal{E} \times \mathcal{C}$.  The set $\mathcal{E}$ ($\vert \mathcal{E} \vert \geq 1$) has to include typical environmental conditions of the system. For systems embedded in an external environment with seasonal variations, this ensemble has to include standard external parameter sequence values (e.g., temperature, irradiance, pressure, wind speed)  encountered over different seasons. The set $\mathcal{C}$  ($\vert \mathcal{C} \vert \geq 1$) of typical command values has to include values that span the allowed input range and that are expected to modulate important variations in the system response.

During retraining,  the $L^{1}$-norm $ \| \hat{o}_{\text{simu}} -  \hat{o}_{\text{retrained}}  \|_1$ between the initial reduced model trained on the simulation and the actual reduced model being retrained on the historical data set is frequently evaluated on $\mathcal{E} \times \mathcal{C}$. A set of parameters are introduced to control the retraining: $\delta_{M}>0$, that quantifies the maximal allowed discrepancy between the two models and is used to stop the retraining, $n_{\rm max}$, that is used to limit the number of retraining epochs on historical data, and $p\ll n_{\rm{max}}$, that is used to set the evaluation period on  $\mathcal{E} \times \mathcal{C}$. The choice of values for $\delta_{M}$ and  $n_{\rm max}$ depends on the system as well as the size and variability of the historical dataset. Typical values of these parameters are given in Sections \ref{pendulum_case} and \ref{building} for the presented test cases. One of the core ideas of our method is not to create a single very accurate reduced model of the real system by fine-tuning $\delta_{M}$ and  $n_{\rm max}$, but rather to use these parameters to create an ensemble of reduced models, that can be used to build a robust controller (see Section \ref{optimal_training}). 

To enhance the reduced models variability, we propose to use two retraining algorithms, both based on the previously-explained stopping rules. In the first algorithm, Algorithm \ref{full},  the weights $\theta$ of  the entire network architecture (encoder, decoder, MLP) are retrained on real data. Such full retraining procedure may still overfit  historical data in cases where very little variability is observed in the control variables and/or if the historical data set is small. To overcome this issue and get system representations that are closer to the simulation, we use a second algorithm, Algorithm \ref{partial}, inspired by recent works on model fine tuning for natural language classification tasks \cite{howar}. In  Algorithm \ref{partial}, only the MLP layers weights are updated by retraining on historical data. Moreover, the training is done in a gradual way, from the output layer to the internal layers, with a learning rate exhibiting a triangular shape, in a similar fashion to \cite{howar}. The learning rate peaks after a few iterations and then exhibits a slow decrease. 

The rationale behind the second retraining technique is to get models that converge fast to a region in the parameter space (fast increasing slope of the learning rate) and then to refine them around this region (slow decreasing slope after the peak). It comes at the expense of introducing additional hyperparameters: the decay rate $\mu$, quantifying the decrease of the learning rate as a function of the depth of the layers, as well as the shape coefficients of the triangular learning rate (see Algorithm \ref{partial}). We chose here $\mu = 2.6$, as proposed in \cite{howar} as well as similar parameters for the triangular shape ratios as the ones proposed in \cite{howar} (see Appendix \ref{hyperparameters}). In Algorithm \ref{partial}, we denote by $\lambda(n)$ the learning rate at step $n$.

% \begin{figure}[th]
%\vskip 0.2in
%\begin{center}
%\centerline{\includegraphics[scale=0.48]{lr}}
%\caption{Learning rate profile for partial retraining.}
%\label{lr1}
%\end{center}
%\vskip -0.2in
%\end{figure}

\begin{algorithm}[t]
   \caption{Full Retraining}
   \label{full}
\begin{algorithmic}
   \STATE Choose  batch size $m$, maximal iterations $n_{\rm max}$, $n_{\rm{iter}} = 0$
   \STATE  Load network weights $\theta =\theta_{\rm{sim}} $ (trained on simulation)
   \STATE Choose $p \ll n_{\rm{max}}$ and  $\delta_{M}>0$
   \REPEAT
   \STATE Sample minibatch of historical  data $\mathcal{Z}$ of size $m$
  \STATE Update   $\theta$ by stochastic gradient descent on $\mathcal{Z}$
   \IF{$ n_{\rm{iter}} \equiv 0 \textrm{ (mod p)}$}
   \STATE Evaluate $\delta:=   \| \hat{o}_{\rm{simu}} -  \hat{o}_{\rm{retrained}}  \|_1$  on  $\mathcal{E} \times \mathcal{C}$
   \STATE $ n_{\rm{iter}} \leftarrow n_{\rm{iter}}  + 1$
   \ENDIF
   \UNTIL{$\delta > \delta_{M}$  \OR $ n_{\rm{iter}}>n_{\rm{max}}$  }
\end{algorithmic}
\end{algorithm}

\begin{algorithm}[h]
   \caption{ Fine Retraining}
   \label{partial}
\begin{algorithmic}
   \STATE Choose  batch size $m$, maximal iterations $n_{\rm{max}}$, learning rate $\lambda = \lambda_0$
   \STATE  Load network weights $\theta =  (\tilde{\theta}_{\tiny enc}, \tilde{\theta}_{\tiny dec }, (\tilde{\theta}_{l})_{l \in L }) = \theta_{\rm {sim}} $  %(trained on simulation)
   \STATE Choose $p \ll n_{max}$,  $\delta_{M}>0$ and $\mu \in (0,1)$.
    \FOR{  $ l \in \{1,...,L\}$}
    \STATE  $n_{\rm{iter}} = 0$
   \REPEAT
   \STATE Sample minibatch of data $\mathcal{Z}$ of size $m$
  \STATE  Update   $\tilde{\theta}_{l} $ with stochastic gradient descent and learning rate $\lambda(n_{\rm{iter}}) \cdot \mu^{l}$
   \IF{$n_{\rm{iter}} \equiv 0 \textrm{ (mod p)}$}
   \STATE Evaluate $\delta =    \| \hat{o}_{\theta_{\rm{sim}}} - \hat{o}_{\tilde{\theta}_{\tiny \rm{enc}}, \tilde{\theta}_{\tiny \rm{dec}}, (\tilde{\theta}_{l})_{l \in L }} \|_{1} $ on $\mathcal{E} \times \mathcal{C}$
   \STATE $ n_{\rm{iter}}\leftarrow n_{\rm{iter}}  + 1$
   \ENDIF
   \UNTIL{$\delta > \delta_{M} $  \OR $ n_{\rm{iter}}>n_{\rm{max}}$  }
   \ENDFOR
\end{algorithmic}
\end{algorithm}

\subsection{Step 3: Training of optimal controller}
\label{optimal_training}

The procedure described in \ref{nn} makes it easy to produce an ensemble of reduced models of the real dynamic system. These reduced models can be interpreted as interpolations between the real system and the simplified simulated system obtained after the first step and described in Section \ref{simple_sim}. Such models could be directly used for MPC by building an optimizer on top of one or more models, depending whether deterministic or stochastic optimization methods are chosen.  We present here an alternative path that can take full advantage of the ensemble of reduced models constructed, to  get a robust controller, that is expected to be ready for real deployment after an offline training phase.  We propose to use an RL method, Proximal Policy Optimization (PPO) \cite{schulman2017}, which is an actor-critic RL algorithm that has shown great potential on control tasks \cite{ppo_18}. The main features of this algorithm are as follows: 
\begin{itemize}
\item It is based on reinforcement learning i.e., on interactions between an actor, represented by a policy map $\pi$ from the observation space to the action space, and its environment (here the reduced models). The actor tries to maximize its expected discounted sum of rewards $r$ i.e., the policy $\pi$  is so as to maximize
\begin{equation}
V_{\pi}(s) = \mathbb{E}_{\pi} \left( \sum_{l=0}^{\infty}  r(s_{l+1}) \gamma^{l}  \mid  s_0 =s \right)
\end{equation}
for every state $s$ of the environment. Here $\gamma \in (0,1)$ is the discount factor. In most real problems,  the reward function $r$ has to be tuned by hand such as to reach a given final  objective. 
\item It is an actor-critics framework, the actor policy $\pi$ is represented by a neural network,  and the value function $V_{\pi}$ that is used to estimate the advantage function $A_{\pi}$ is represented  by another neural network.  The advantage function corresponds to the mean ``advantage'' of taking a given action $a$ when the environment is in the state $s$; see \cite{schulman2015trust} for more details.
\item It is an on-policy algorithm i.e., relies on Monte Carlo estimates to compute the advantage function. To gain in sample efficiency, it uses importance sampling.
\item It is based on a trust-region optimization scheme: it improves the policy by looking for a maximizer of the estimated advantage  in a small region concentrated on the actual policy,  at every optimization step.
\item It provides a theoretical improvement guarantee at every iteration step: 
\begin{equation}
\mathbb{E}_{\rho_0} V_{\pi_{n+1}}  \geq  \mathbb{E}_{\rho_0} V_{\pi_{n}},
\end{equation}
where $\mathbb{E}_{\rho_0} V_{\pi} = \sum_s \rho_0(s) V_{\pi}(s)$ and $\rho_0$ is the initial environment state distribution. However, this theoretical guarantee vanishes as soon as we employ function approximators like neural networks for the policy map and the advantage function.
\end{itemize}
In RL, constraints (e.g.,  minimal temperature constraints, maximal torque speed) are usually integrated as soft constraints in the reward function and we adopt the same strategy here. If $E(s) \geq 0$ is the objective to maximize under the constraint  $f(s,a) = 0$ , the reward $r(s,a)$ is usually shaped as \begin{equation}
r(s,a) = E(s) -  \upsilon  f^2(s,a),
\end{equation}
where $ \upsilon  >0$ has to be tuned to ensure that violations of  $f(s,a)=0$ dominate largely the objective $E(s)$ in state $s$ where they are  unacceptable. Moreover, as in recent works, policy and advantage functions are represented with neural networks (either MLP or LSTM networks).

The main new ingredient of our method is to use the set of reduced models created at Step 2 to train the controller in a distributed manner similar to \cite{heess2017}, which shows improved policies with distributed PPO. We use the reduced models' diversity in the same way as dynamics randomization is used in \cite{peng2018sim} to create robust controllers with reinforcement learning. We also introduce refined exploration noise approaches for continuous and discrete variables. This strategy led to more stable policies in the system test case presented in Section \ref{building}. In the following, we detail the distributed training and noise exploration strategies.

\subsubsection*{Distributed training}
 
\begin{figure}[ht]
\begin{center}
\centerline{\includegraphics[scale=2.5]{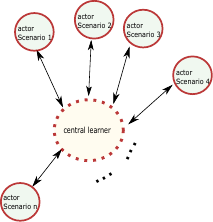}}
\caption{Distributed framework for controller training. The different reduced models are used in parallel at the slave nodes level.}
\label{dist_learning}
\end{center}
\end{figure}

Let $N$ be the number of reduced models identified in step 2 (Section \ref{nn}). We use $N-K$, where $K \ll N$, reduced models to train the controller in a distributed manner. The remaining $K$ models are kept aside for evaluating the performance, reliability and robustness  of the controller after RL  training. At each actor node level in Figure \ref{dist_learning}, one of these $N-K$ models are used to represent the real system dynamics. A copy of the actor is acting on each of these models individually. Data from the actor copies are gathered at a central level, where policy and  value function network weights are updated  following the PPO algorithm, see  Algorithm 1 in \cite{schulman2017}: Each decentralized actor copy carries out $T$ rollouts with its attached environment model and rewards and generalized advantage are estimated for the actor copy at each timestep using Eq. (11)-(12) from \cite{schulman2017}.  The rollouts vectors are then gathered at the central level to form $(N-K)T$ timesteps of data with advantage $\hat{A}_t$ estimated at each timestep t. Finally, minibatches of size $M \leq (N-K)T$ are randomly sampled from the rollout vectors. The surrogate loss is estimated over each minibatch by averaging the advantage estimates with importance sampling weights coming from the central learner policy (Eq. 7 in \cite{schulman2017}). Surrogate loss and value function loss are then minimized over the central learner parameters using stochastic gradient descent. Minibatch sampling  followed by gradient descent is repeated $L$ times at the central level.  We represent this centralized update with a ``central learner''. After every central update, the weights of the actors at each node are updated with the new weights.   Using this distributed learning technique not only improves the time needed to train the controllers, it also allows them to be trained in a robust manner as every actor copy is run on a different reduced model of the system under different external conditions.

\subsubsection*{Noise and exploration strategy}
Policies for actions that take continuous values are parametrized by multivariate Gaussian  distributions  $\pi_g \sim \mathcal{N}(\mu(s), \sigma(s))$ with mean  $\mu(s)$ and covariance matrix $\sigma(s)>0$. The policy neural network trained by PPO predicts the tuple $(\mu(s),\sigma(s))$ for every state $s$ by projecting the output of the last hidden layer of the MLP network (or the LSTM output $h$ if a LSTM network is used)  to the required control dimensions with a linear transformation.   For actions that are constrained to lie in bounded sets,  a tangent hyperbolic transformation  is applied to map the action vector to a  bounded set. In particular, if $ a \in [\alpha,\beta]^n$, then the agent actions $a$ are multivariate random variables defined by 
\begin{equation}
a(s) := \frac{\beta-\alpha}{2} \tanh(\pi_g(s)) + \frac{\alpha+\beta}{2}
\end{equation}
Using Gaussian distribution ensures enough exploration during the training phase. Once the policy is evaluated on a particular set after training,  the output $\sigma(s)$ of the neural network is ignored and one considers the deterministic policy
\begin{equation}
\pi_{\rm{det}}(s) = \frac{\beta-\alpha}{2} \tanh(\mu(s)) + \frac{\alpha+\beta}{2}.
\end{equation}

For policies with discrete actions, a different strategy is used for noise exploration. If $a$ can take $k$-discrete values, the neural network $\pi_{d}$ maps any environment state $s$ to $\mathbb{R}^k$.  Let $Y$ be a 2-valued random variable following the binomial distribution with success probability $p$ and $U$ a real random variable in $[0,1]$ following the uniform distribution on $[0,1]$. As the response of real systems may exhibit latency or be slower than the controller,  we do not necessarily change exploration noise at each timestep as done in standard approaches. Instead, we use the following refined rules to sample actions at time $t$ during training:
\begin{itemize}
\item Store the noise $n(t-1)$ at timestep $t-1$, $n(0)$ being equal to $0$.
\item Compute $\pi_{d}(s,t) \in \mathbb{R}^k $ with the neural network. 
\item Sample $Y$ once and get $Y(t)$. Sample $U$ once and get $u(t)$. Update the noise $n(t)$ via
\begin{equation}
n(t) = n(t-1) *(1-Y(t)) +   Y(t) \cdot U(t)
\end{equation}
\item Take action \begin{equation} a(t) = \textrm{argmax}[ \log( -\log(n(t))) + \pi_{d}(s,t)]. \end{equation}
\end{itemize}
 By evaluating, we take the deterministic policy  \begin{equation} \textrm{argmax}[ \pi_{d}(s,t)]. \end{equation}

In both discrete and continuous cases,  we add observation noise  to prevent controllers from overfitting flaw patterns of the reduced models used to represent the real system. We employ here an  Uhlenbeck-Bernstein process to perturb the outputs of the reduced models. Using a process with drift instead of pure Gaussian noise forces the controllers to rely on more observation variables for control and to react to stochastic variations in the system response that may not have been captured by the reduced models but have a middle to long-term influence.

\section{Benchmark of the system identification process}
\label{pendulum_case}
  In order to illustrate the generalization capabilities of our system identification process (steps 1 and 2, Section \ref{sec2}) to new input commands, we design a system identification experiment with a known simple system and where the historical training data have been generated using restricted input signals.
% and evaluated on unseen excitation signals <- what has been evaluated?
We compare our method against other methods, especially the established numerical subspace state space system identification (N4SID) \cite{van2012subspace}. We choose the pendulum with external torque as the non-linear system to identify. The equations of motion of the pendulum are given by
\begin{equation}
\label{nonlinear}
\ddot \theta(t) + \frac{g}{l}\sin(\theta(t)) = -\frac{T(t)}{ml^2},
\end{equation}
where $T(t)$ is the external torque applied to the pendulum at time $t$, $l$ the pendulum length, $m$ its mass, and $g=10$ $ms^{-2}$ the gravitational constant. We set $(l,m)=(1,1)$  a.u. 

\begin{figure}[ht]
\begin{center}
\centerline{\includegraphics[scale=0.8]{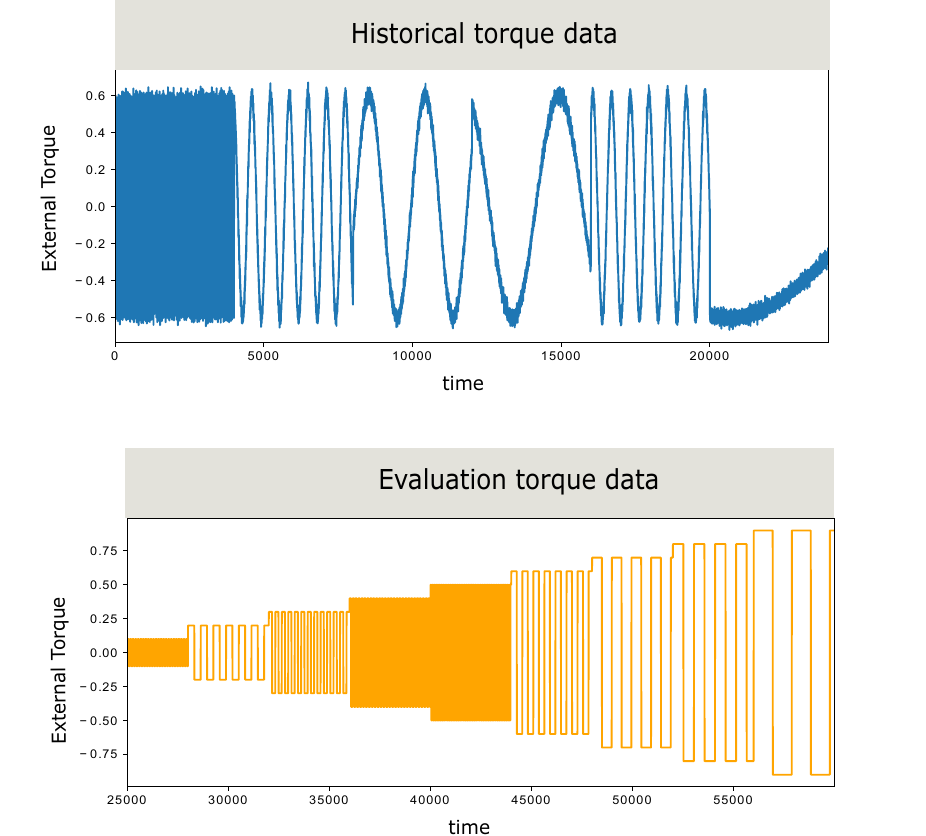}}
\caption{Historical torque (top) and evaluation torque (bottom) signals $T(t)$.}
\label{made_up}
\end{center}
\end{figure}

Our identification experiment presents the issues for which our method has been designed. Firstly, the historical data for training is limited in size (25K samples) and generated under a restricted torque input signal with sinusoidal shape and a fixed amplitude (see Figure \ref{made_up} top panel). Secondly, we also assume that the real dynamic model given in \eqref{nonlinear} is only partially known, and hence we construct a semi-physical simulation that only captures some of its properties. 

The identified models are tested on an evaluation data set generated with the true dynamics described by \eqref{nonlinear} and excited with square torque signals with varying amplitude range (see Figure \ref{made_up} bottom panel). To make the task a little bit more difficult and to check if the mathematical constraints are well captured by the reduced model, we did not try to predict $\theta$, but rather the normalized observations tuple $(x,y) = (\sin(\theta),\cos(\theta)) \in[-1,1] ^2$. In the following we detail step 1 and 2 for the experiment as well as the obtained results. 

\subsection{Step 1: Semi-physical simulation}
\label{semi_phys}
As approximate simulation of the real system, we consider the linearized equations
\begin{equation}
\label{linear}
\ddot \theta(t) + \frac{g}{l'} \theta(t) = -\frac{T(t)}{m'l'^2}.
\end{equation}
Solutions of  Eq. \eqref{linear} are good approximations of solutions of  \eqref{nonlinear} only if $\theta \ll 1$.  Clearly, \eqref{nonlinear} can be linearized around any point $\theta_0$ and  if $\theta$ starts to deviate from the vertical axis, a new linearized equation could be used to predict its dynamics. However, we would like to check here how a ``wrong'' simulation can be corrected with real collected on-site data with our method, and this is why we chose to use Eq. \eqref{linear} for all values of $\theta$.
To further complicate the identification task, we assumed that $m'\neq m$ and $l'\neq l$ and chose $(l',m')=(1.2,1.5)$ a.u.. This corresponds to the case where only vague estimates of the pendulum mass and length have been provided.
% We then discretized the equations of motions for \eqref{nonlinear} and \eqref{linear}. For the linear case, it yields to
%\begin{equation}
%\begin{split}
%\dot \theta[p+1] &= \dot \theta[p] - \frac{g}{l'} \theta[p] dt - \frac{T[p]}{m'l'^2}dt,\\
%\theta[p+1] &= \theta[p] + \dot \theta[p+1] dt.
%\end{split}
%\end{equation}
Using Eq.\ref{linear}, 250K data samples were generated using input torque signals with random amplitudes (drawn from a uniform distribution in the interval $[-2,2]$ a.u.) and varying frequencies.

\subsection{Step 2: System Identification}
The reduced models have one input $T(t)$ and two outputs $x(t)$ and $y(t)$. They were first trained on the 250K samples generated by the semi-physical model and then retrained on 25K of data  generated by the real system (Eq.  \eqref{nonlinear}) with sinusoidal torque excitations (Figure \ref{made_up} top panel).  Retraining was done with Algorithm \ref{full} for $\delta_M = 0.2$ and varying values of $n_{max}$, the evaluation set $\mathcal{E}\times \mathcal{C}$ being a fixed sequence of length 2K from the 250K linearized samples.

\subsection{System identification results}
 
\begin{figure}[ht]
\begin{center}
\centerline{\includegraphics[scale=0.4]{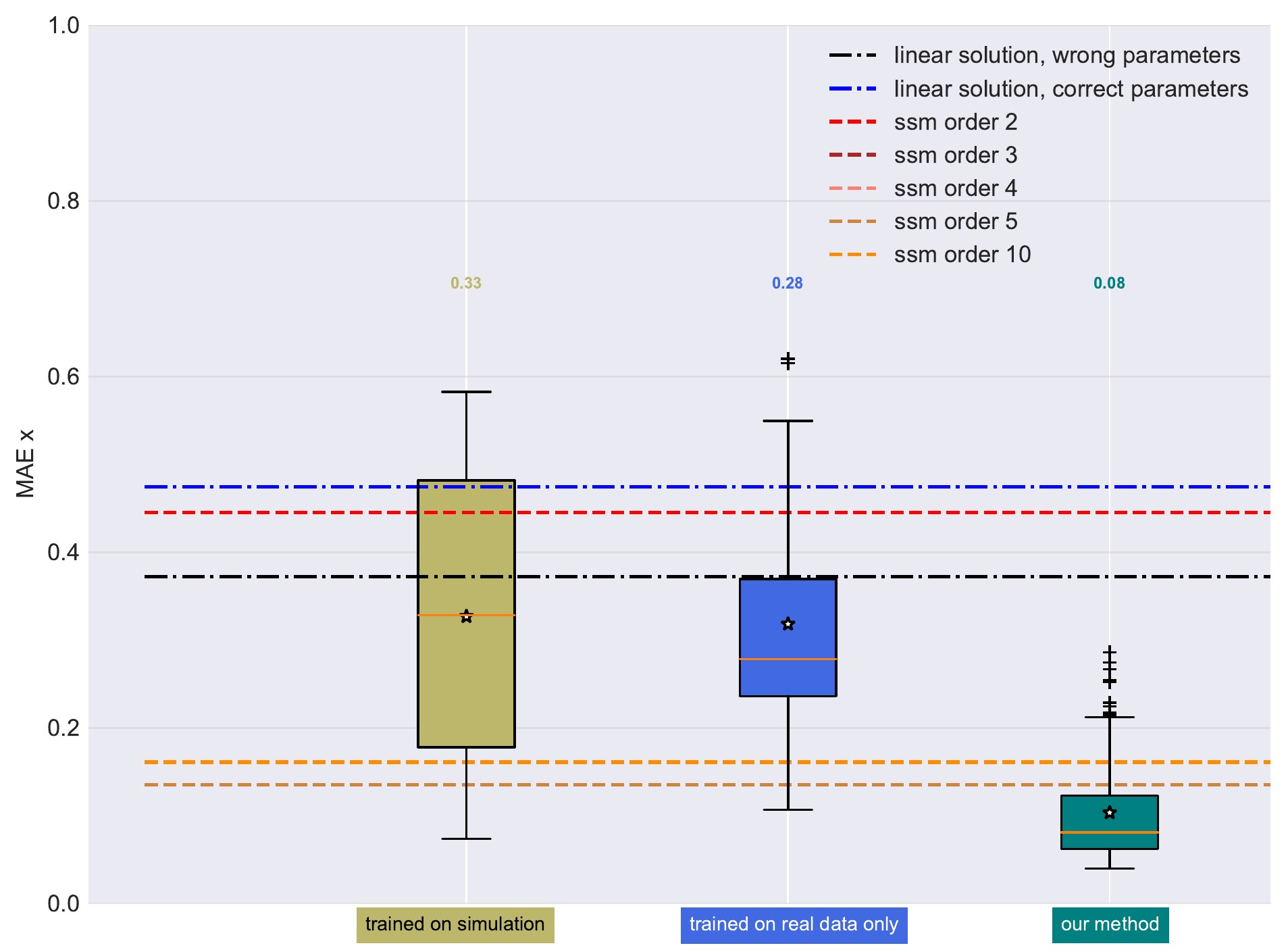}}
\caption{Mean absolute error for $x-$coordinate. Models obtained with our method  outperform linear state space models identified with N4SID (ssm 2 to 10). Evaluation carried out on 30K steps, with unrolling of size 1000.}
\label{res1_1}
\end{center}
\end{figure}

\begin{figure}[ht]
\begin{center}
\centerline{\includegraphics[scale=0.4]{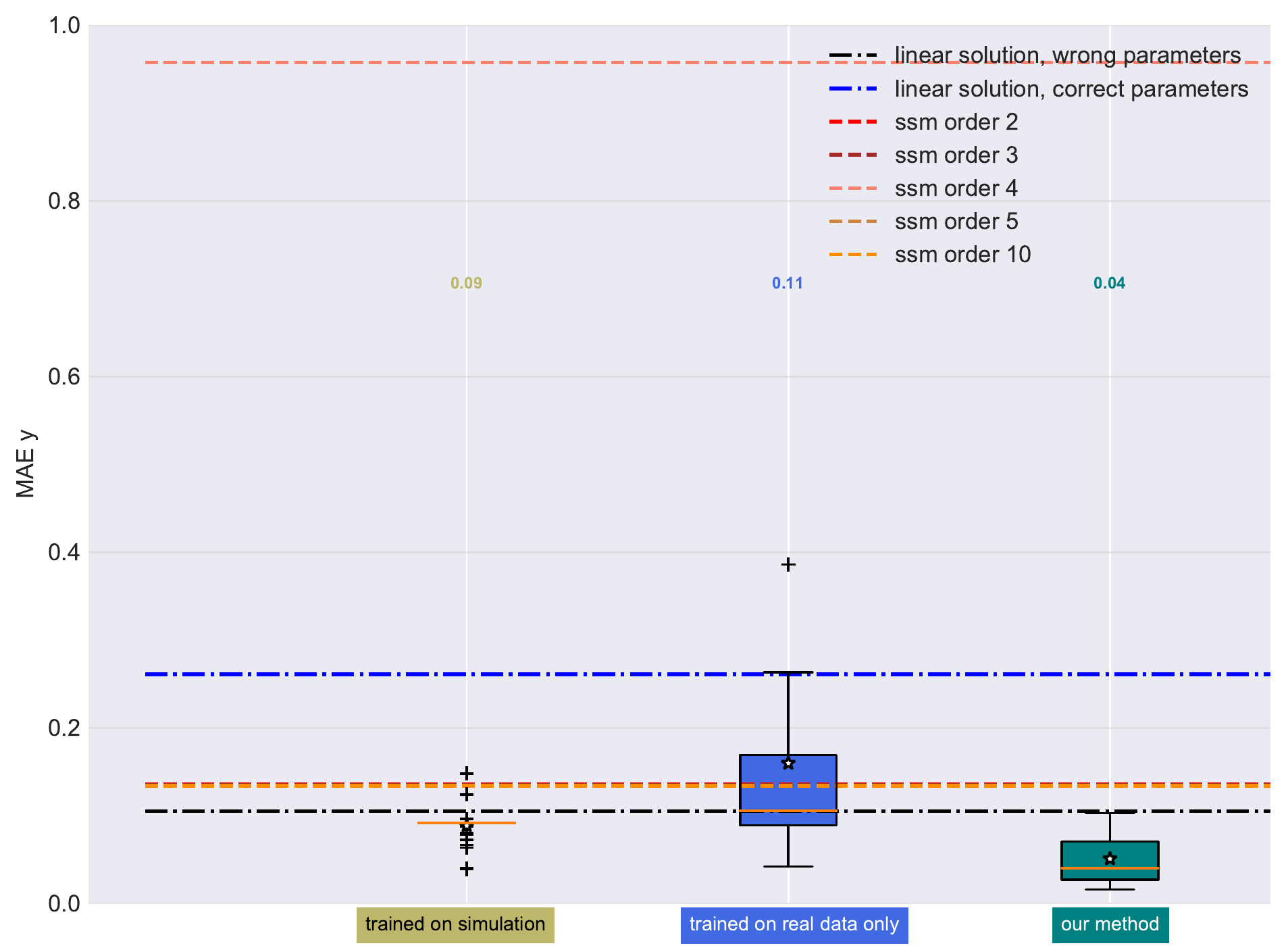}}
\caption{Mean absolute error for $y-$coordinate. Models obtained with our method clearly outperform other approaches and linear state space models identified with N4SID  (ssm 2 to 10). Evaluation carried out on 30K steps, with unrolling of size 1000.}
\label{res1_2}
\end{center}
\end{figure}

\begin{figure}[ht]
\begin{center}
\centerline{\includegraphics[scale=0.4]{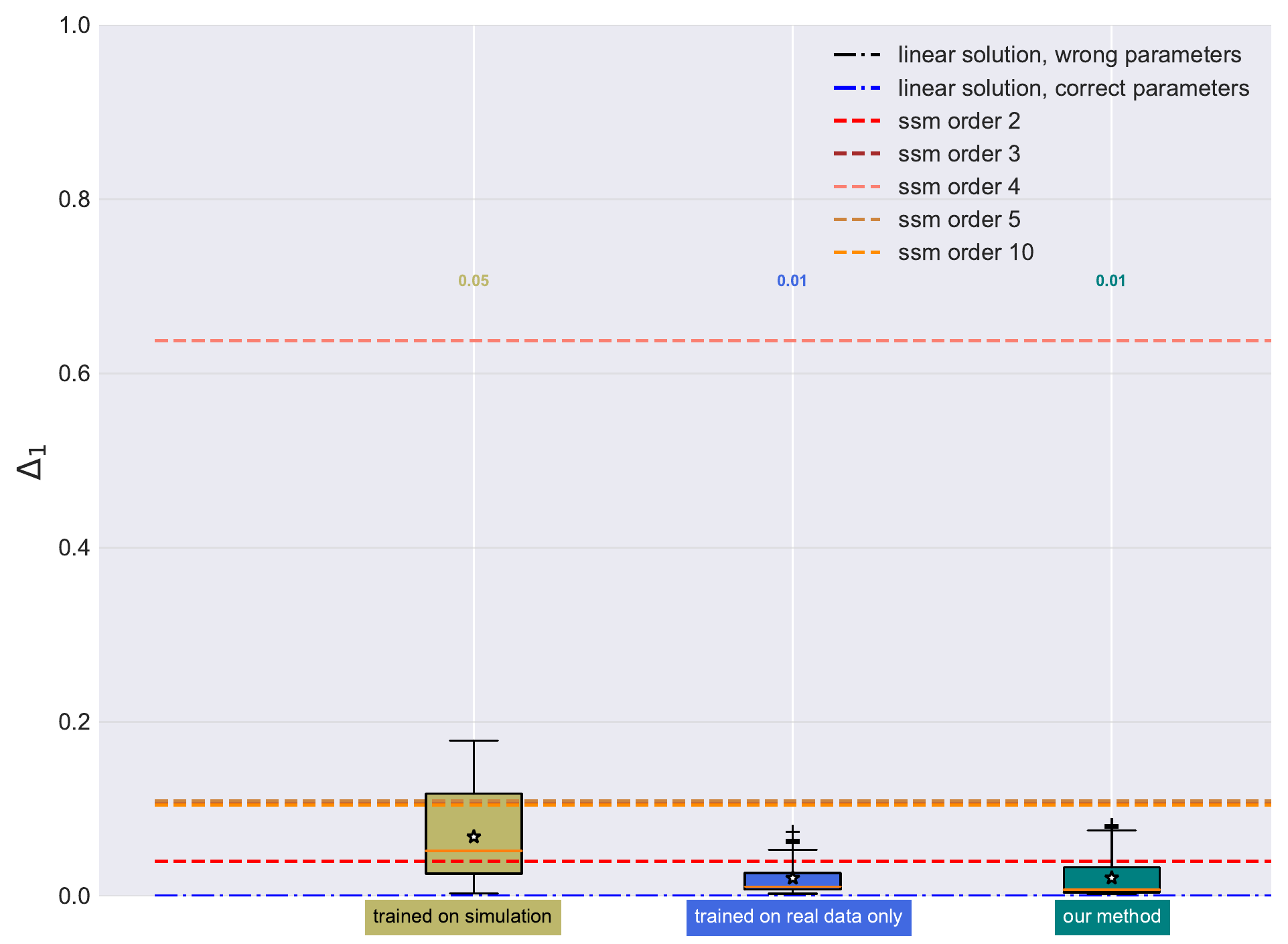}}
\caption{$\Delta_1 =\sqrt{\hat{x}^{2}(t) + \hat{y}^{2}(t)} -1$  is an indication of model physical discrepancy.  Models obtained with our method outperform linear state space models identified with N4SID  (ssm 2 to 10).}
\label{res1_3}
\end{center}
\end{figure}

The generalization capability to new input commands is assessed by evaluating the accuracy of reduced models predictions on a set of 30K points of real data generated with square input torque signals (Figure \ref{made_up} bottom). Mean absolute error (MAE) and the mean deviation $\Delta_1 = \frac{1}{N}\sum_{t \in N} \mid \sqrt{\hat{x}^2(t) + \hat{y}^2(t)} -1\mid$ ($N$ being the number of evaluated samples) are used as metrics and shown in Figures \ref{res1_1} - \ref{res1_3} for the reduced models trained with our method and a list of other models used as benchmark. The benchmark models include: the linear model directly following  \eqref{linear}, both with correct and incorrect mass and length; linear state space models identified on historical data with the N4SID method (\cite{van2012subspace}) and varying hidden dimension of the state space variable (from 2 to 10) . We also compare against LSTM encoder-decoder models trained only on simulated data and trained only on real data with sinusoidal excitations. To address stochastic effects in our training method, we report results for 150 reduced models obtained with varying parameters $n_{\rm max}$.  

Reduced models trained with our method outperform all other models for all metrics, which shows the generalization capabilities of the proposed method.

\section{Benchmark of the full method}
\label{building}
We evaluated the full pipeline of our method by applying it to control a large office building facility. We compare controller performance with rule based controllers under a representative set of external condition and show that the purposed method outperforms the latter.  

The building is a 3-floor modern office building with 100 rooms. It has two central ventilation units feeding each approximately half of the rooms. It has a global cooling and a global heating fluid based central unit. Rooms have different types of inputs. They are equipped with a thermostat that controls valve opening for heating and/or cooling. In most of the rooms, valve opening is linked to a changeover because the same pipes are used both for heating and cooling. In the rest of the rooms, only cooling or heating is available.
\begin{figure}[ht]
\begin{center}
\includegraphics[scale=1.2]{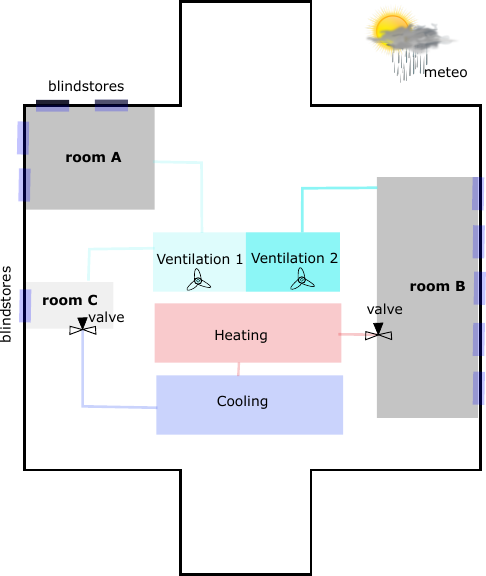}
\end{center}
\caption{Schematic view of the building showing three possible room configurations.}
\label{EPFL}
\end{figure}
Blinds opening and closing are automatically controlled with fixed closing rules during midday break, but the user can manually take over blinds control when they wish. Heating setpoints for global heating follows a  heating curve depending on the season and the mean outdoor temperature.  As the building envelope is made with a lot of glass,  solar radiance exerts a high influence on the room temperature, and the building experiences relatively high temperatures in south-exposed rooms during summer times. Moreover, the ventilation units of the building have been undersized. A schematic view of the building control, input and outputs, are shown in Figure \ref{EPFL}.  A certain number of unknown factors impact building models accuracy. In particular, room occupancy and door openings and closings are not measured. Moreover, heating, cooling and ventilation temperatures are only measured at the output of the central units.  These uncertainties are affecting models accuracy. However, our method to train the controllers is able to tackle such issues with the addition of noise and model randomization (see Sections \ref{optimal_training} and \ref{distrRL}).

\subsection{Step 1: Semi-physical simulation}
\label{simu}
Within the scope of the european project THERMOSS, in which part of this work was carried out, a simple python-based building and district simulator, DIMOSIM (DIstrict  MOdeller  and  SIMulator,  \cite{perez2015thermal}) was provided by  France's CSTB \cite{CSTB}. The thermal models used in DIMOSIM are based on RC modeling i.e.,  on linear models for evolution of air and internal layers temperatures. The simulator provides ray-tracing algorithm for solar gain calculations with masks. We used that simple framework to construct blocks of the building. 
To simplify the modeling work, we defined room classes $\mathcal{C}_i$, where every class $\mathcal{C}_i$ is defined by a tuple $(I_i, F_i,E_i,S_i)$, with $I_i$ the list of inputs/outputs for the room,  $F_i$ the floor number (from 0 to 2), $E_i$ the room orientation (south, north, west, east, south-east, south-west, ...), $S_i$ the size of the room (big, medium, small).

With this class definition and taking into account the building properties, we obtained thirty different classes. Every room belongs to one of these classes and we assumed that the ground thermal dynamics is the same for all rooms in the class, making it sufficient to only simulate one member of the class. We automatized as much as possible the room construction, constructing all properties of the room from a single text file summarizing inputs and outputs for all rooms,  building footprints, height of the building, window area per room class, and best-guessed conductivity, density and heat capacity of the building envelope. We did not try to fine-tune those parameters.

Room classes were simulated once at a time, with randomly chosen input setpoints and weather conditions from eleven locations around the world. To simplify the modeling and avoid using advanced equipment classes, we  used the ventilation class to drive the thermal properties in the room with the input commands, again, in a purely linear manner: let $c_i$ be a given representative of the room class  $\mathcal{C}_i$, with valve opening control $O(t)$, central fluid supply temperature $T_{\rm{flow}}(t)$, air change rate $C_r(t)$,  and air temperature flow $T_{\rm{air}}(t)$. Then, at time-step $t$,  the effective room ventilation temperature  is:
\begin{equation}
\label{eq1}
T_{\rm{eff}}(t)  = \frac{T_{\rm{air}}(t) \cdot C_r(t) + \alpha \cdot T_{\rm{flow}}(t) \cdot O(t)}{ C_r(t) + \alpha  \cdot O(t)  },
\end{equation}
where $\alpha$ is a coefficient that can be tuned to better fit the real room model. We found out that $\alpha = 0.6$ yields credible room temperature models for most of the rooms and we did not try to fine-tune it room by room. The timestep for the simulation was set to ten minutes and sensor values that have different sampling frequencies were interpolated on a ten minutes basis.  Equation \ref{eq1} assumes that the air blown in the room and the room air cooled/heated by the source reach equilibrium after 10 minutes. Moreover, \eqref{eq1} ignores losses in the ventilation and heating/cooling system, as $T_{\rm{air}}$ and $T_{\rm{flow}}$ are only measured at the central level.  The inputs commands and setpoints for the temperatures and openings (blinds, valves) for training were chosen at random and with random frequencies to avoid overfitting a fixed pattern when we trained the reduced model on the outputs.

\subsection{Step 2: System identification}
\label{sysid}
We modeled every room with an encoder-decoder architecture, as explained in Section \ref{sec2}, the output $o$ being the room temperature, and the inputs (external and setpoints) being the ones presented in Table \ref{table_building_in}. 
\begin{table}[H]
\caption{Global and local inputs used to model room temperatures}
\label{table_building_in}
\vskip 0.15in
\begin{center}
\begin{small}
\begin{sc}
\begin{tabular}{ll}
\toprule
\textbf{Global inputs}                                       & \textbf{Local Inputs}   \\
\midrule
Actual meteorological values & Blind positions  \\
Heating fluid temperature                    & Valve positions       \\
Cooling fluid temperature                    &                                  \\
Ventilation supply temperature              &          \\          
\bottomrule              
\end{tabular}
\end{sc}
\end{small}
\end{center}
\end{table}

The total building graph is the collection of the single room graphs, with shared placeholders for the global input commands (ventilation, cooling/heating) and the external parameters (exterior temperature, irradiance). Training first on the simulated room classes, the squared-loss reached after a few thousand timesteps values around 0.27 \degree C, which we considered sufficient for our purpose and the high variability of the conditions imposed in the simulation. 

Special care had to be taken when retraining with historical data from the test site as they were obtained under closed loop control with fixed  temperature bounds around 23°C for the rooms.
 
We chose the control set  $\mathcal{E} \times \mathcal{C}$ as follows: $\mathcal{E} $ is a set of three specific days, $\{d_{cool},d_{hot},d_{cloud}\}$, $d_{cool}$ being a very cold day, $d_{hot}$ a very hot day, and  $d_{cloud}$ a cloudy day with temperatures around 15-20 \degree C. For each of these days, we chose five possible scenarios which constitute our command set $\mathcal{C}$ and are given in Appendix \ref{AppC}.
 
We retrained every room on four months of historical data (from February to middle of June 2018) following algorithms \ref{full} and \ref{partial}, with stopping set $\mathcal{E} \times \mathcal{C}$.

\begin{figure}[th]
\begin{center}
\centerline{\includegraphics[scale=0.52]{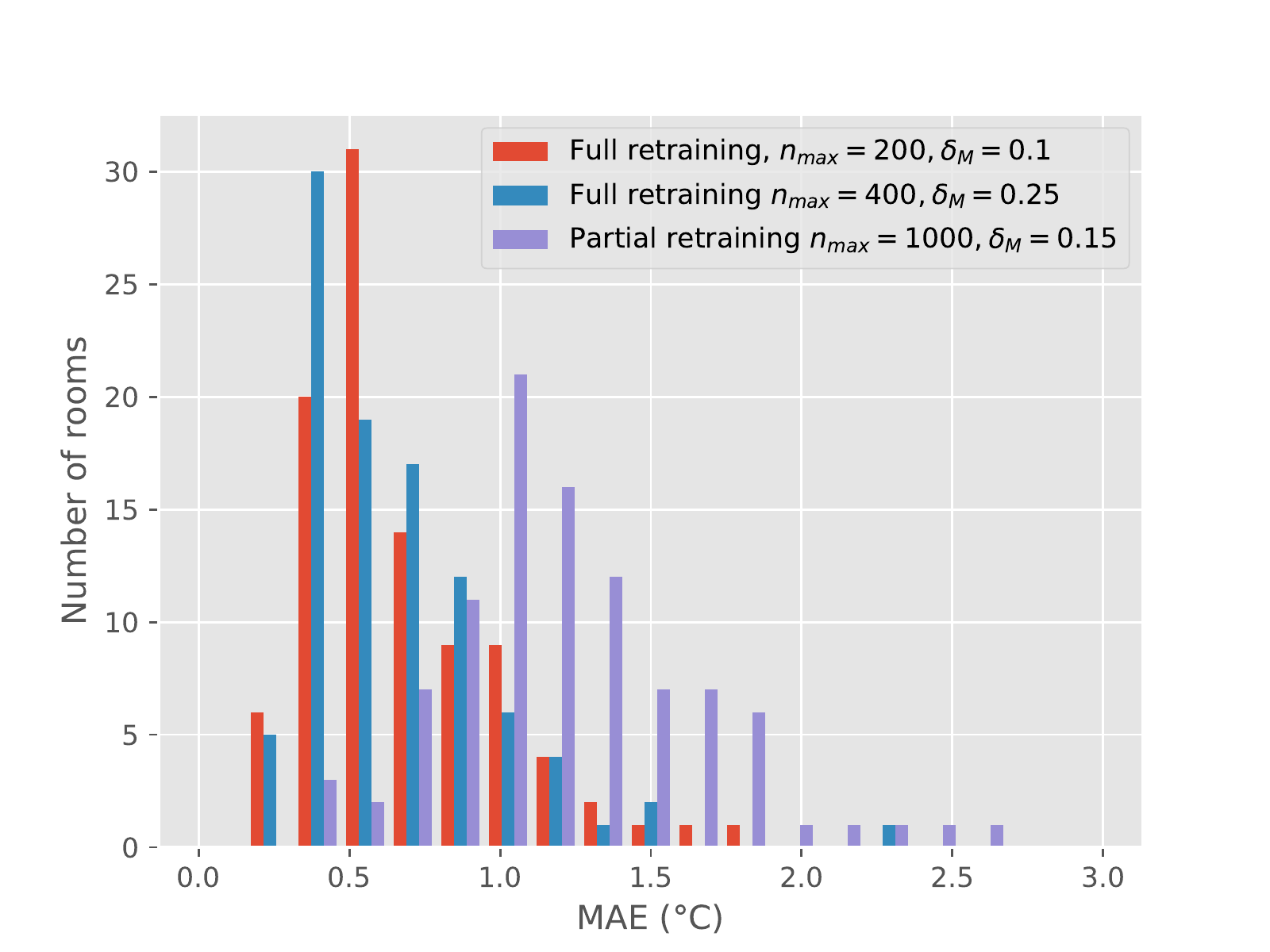}}
\caption{MAE distribution for three building neural models. The MAE  validation set ranges from September 2018 to end of November 2018.}
\label{MAE_dist}
\end{center}
\end{figure}

\begin{figure}[th]
\begin{center}
\centerline{\includegraphics[scale=0.52]{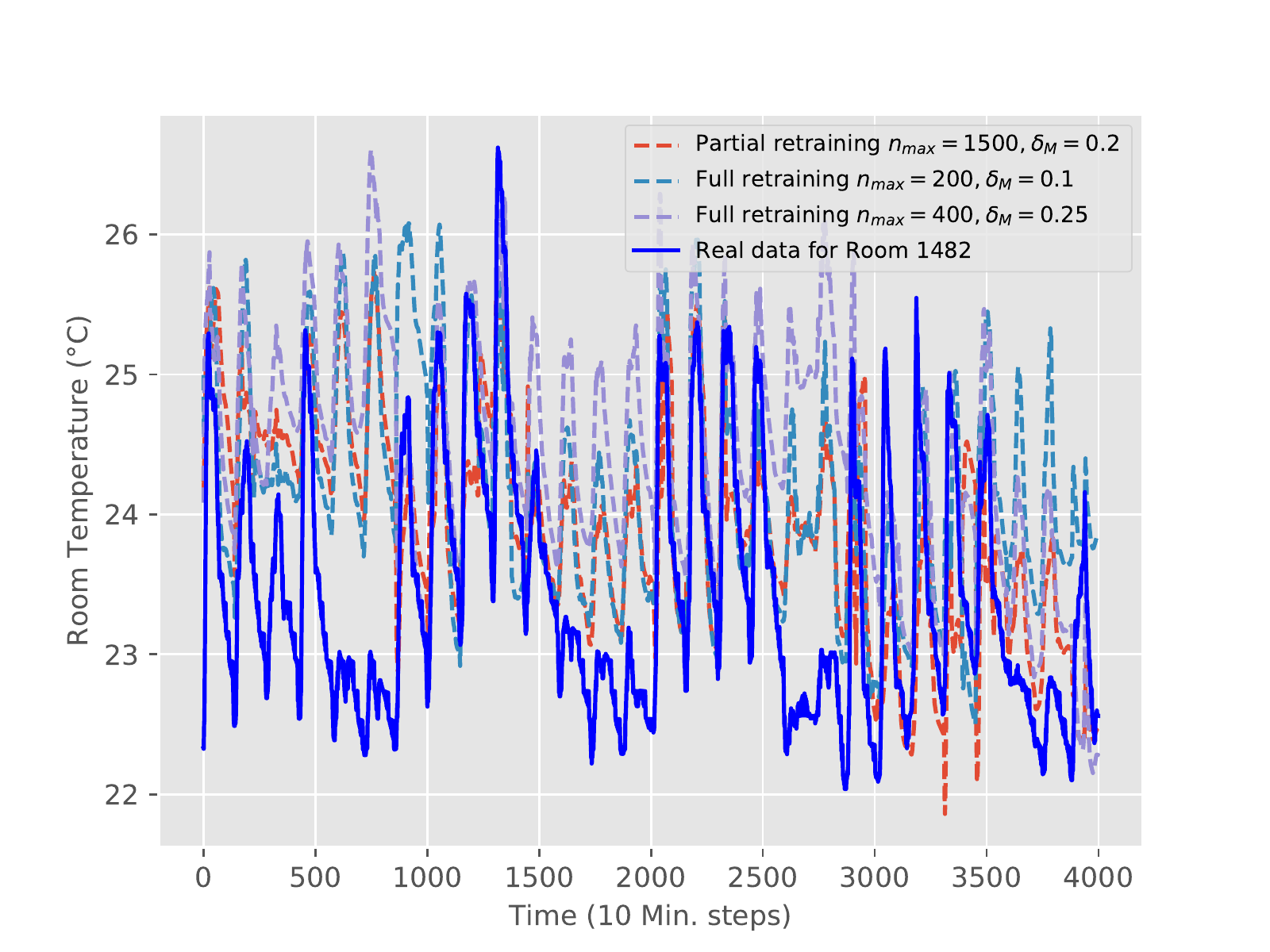}}
\caption{Temperature predictions vs observations for a large room.}
\label{r1}
\end{center}
\end{figure}

The retrained models were evaluated on three months of historical data that were not used for training, from September to November 2018. The predicted data were generated after initializing the encoder with 24 data points (four hours), and unrolling it during  three months, without re-initializing it with observed data. Figure \ref{MAE_dist} shows the histogram of the mean absolute error (MAE) for the room temperatures on real historical data for three different parameters of our training method. Figure \ref{r1} shows an example of the predicted temperature by our reduced models, with the three different training parameters, of a large room, as well as the observed temperature. The results show that the reduced models kept a good stability over the entire unrolling period, which is due to the stability of LSTM networks and their ability to encode and keep long term information in the cell state. The stability of the reduced models is crucial because we use them for modeling the building during entire years without re-initializing them.

Table \ref{Percent-table} reports the MAE percentiles and average for ten models over three months. Its purpose is not to choose the best identified model, but rather to showcase that Algorihms \ref{full} and \ref{partial} are able to create an ensemble of reasonably close physical copies of the real  building,  that interpolate between the simulated building and the real building. Results in both Figure \ref{MAE_dist} and Table \ref{Percent-table} show that models obtained by retraining the entire network (Algorithm \ref{full}) better match observed data on site than models obtained by retraining only the last layers with Algorithm \ref{partial} (with, however, a higher risk of overfit, as discussed in Section \ref{nn}). This is expected, because not all models weights are updated with Algorithm \ref{partial}. Moreover, for Algorithm \ref{partial}, larger values of $n_{\rm \max}$ and $\delta_M$ lead to lower error. All these models (except two of them that are used for validation) will be used in Section \ref{distrRL} to build controllers with distributed reinforcement learning.

\begin{table}[ht]
\caption{MAE percentiles  and mean for ten models over three months. Models in blue were retrained with Algorithm \ref{partial}, i.e., only the MLP layers were updated}
\label{Percent-table}
\begin{center}
\begin{small}
\begin{sc}
\begin{tabular}{lllll}
\toprule
 \textbf{Parameters/MAE (\degree C) }     & \textbf{25\% } &   \textbf{50\%}  &  \textbf{75\%}  &   \textbf{Mean}  \\
\midrule

 $n_{max} = 1000, \delta_{M} = 0.15$  &0.98 & 1.18 & 1.47&1.25  \\

$n_{max} = 1500, \delta_{M} = 0.1$ &  0.93 & 1.16&1.47&1.26    \\

$n_{max} = 1500, \delta_{M} = 0.2$ & 0.82& 1.02&1.33&1.12 \\

$n_{max} = 700, \delta_{M} = 0.1$     &  1.07&1.30&1.70&1.41         \\
 
$n_{max} = 1000, \delta_{M} = 0.15$  &\textbf{0.43}&  0.59&0.81&0.80\\ 
$n_{max} = 200, \delta_{M} = 0.1$ &0.46&0.59&0.83&0.70\\ 
 $n_{max} = 300, \delta_{M} = 0.05$ &0.46&\textbf{0.58}&0.78&\textbf{0.69}\\ 
$n_{max} = 400, \delta_{M} = 0.075$ &0.47&0.59&\textbf{0.76}&0.70\\ 
 $n_{max} = 400, \delta_{M} = 0.25$ &0.44&0.60&0.82&0.77\\ 
 $n_{max} = 500, \delta_{M} = 0.2$ &0.46&0.59&0.80&0.70\\
\bottomrule              
\end{tabular}
\end{sc}
\end{small}
\end{center}
\end{table}

%\begin{figure}[ht]
%\begin{center}
%\centerline{\includegraphics[scale=0.6]{Room_1508}}
%\caption{Temperature predictions vs. observation for a middle size room}
%\label{r2}
%\end{center}
%\vskip -0.2in
%\end{figure}

Energy consumption has not been estimated with the simulation but directly from historical data, taking into account the setpoints for the ventilation and heating/cooling fluid temperatures, as well as the exterior meteorological  parameters. Naive fitting of energy consumption with linear models or non-linear models
% like Support Vector Regression (SVR) using least squares as loss function is quite dangerous in our case as our objective is to optimize the energy consumption and thus change the control setpoints. Indeed, as the system is under closed loop control, the coefficients found by least square estimates come with no physical guarantees. In fact, we observed that linear least square regression without constraints were giving unexpected signs to the coefficients. SVR was leading also to energy models that 
led to spurious behavior once we tried to optimize them with the controllers: ventilation units and  heating/cooling units run in parallel in close loop control and let the models favor non-physical coefficients. We alleviated this problem by taking a rectified linear unit of a linear model with signs constraints on the coefficients (in heating mode, more heat means more energy consumption (and similarly for the temperature of the air blown by the ventilation system) and by imposing the two ventilation units to have the same coefficients in the linear model. Evaluation on unseen data during training gave an accuracy around $80 \%$. 

%A comparison between predicted and observed heating energy consumption is shown in Figure \ref{r2}. 
%\begin{figure}[ht]
%\vskip 0.2in
%\begin{center}
%\centerline{\includegraphics[scale=0.52]{heat_energy}}
%\caption{Heating energy consumption: prediction vs. observation over 14 evaluation days. The model is not able to reproduce the very sharp peaks when the heating is turned on.
%}
%\label{r2}
%\end{center}
%\vskip -0.2in
%\end{figure}

\subsection{Step 3: Training of optimal controllers}
\label{distrRL}

%\noindent \textbf{Controlled variables, policy choices.}
\subsubsection*{Control variables and policy choices}
We trained 100 local room controllers with policy  $(\pi_i)_{i\in \textrm{Rooms}}$ and one global building controller with policy $\pi_g$. The local room controllers  are controlling the blinds openings (the valve control logic cannot be changed in practice) in every room. The global building agent $\pi_g$ is controlling the heating temperature supply, the cooling temperature supply and the ventilation temperature supplies (for the two units).  The inputs and outputs for the global policy are summarized in Table \ref{gp_inout}.   While it would have been possible to train a single controller for all the rooms and the technical systems at once, this path was not taken here. Indeed, such task is not only more difficult because of the large degrees of freedom of the optimization process (more than 100  simultaneous control setpoints to optimize), but it is also less practical and robust in a real application since any room change would require retraining of the entire controller, which is not advantageous for any real building control system.  
\begin{table}[h]
\caption{Global policy inputs and outputs}
\label{gp_inout}
%\vskip 0.15in
\begin{center}
\begin{small}
\begin{sc}
\begin{tabular}{ll}
\toprule
 \textbf{Inputs for $\pi_g$}     & \textbf{Outputs for $\pi_g$}   \\
\midrule
Mean floor temp. & Heating temp.\\
 Mean building temp. &  Cooling temp.       \\
Time                 &  Ventilation temp. (bloc 1\&2)  \\
Meteo parameters            &          \\
Seasonal Ext. temp.  &\\          
\bottomrule              
\end{tabular}
\end{sc}
\end{small}
\end{center}
\end{table}

To ensure that the actions with distribution $\pi_g$ take values in bounded intervals $I = I_1 \times \ldots \times I_4$, we use the $\tanh$ transformation to map the action vector to a  bounded set as described in Section \ref{optimal_training}. The inputs and outputs for the local room policies are summarized in Table \ref{lrp_inout}.
\begin{table}[h]
\caption{Local room policy inputs and outputs}
\label{lrp_inout}
\vskip 0.15in
\begin{center}
\begin{small}
\begin{sc}
\begin{tabular}{ll}
\toprule
 \textbf{Inputs for $\pi_i$}     & \textbf{Outputs for $\pi_i$}   \\
\midrule
Room temp.& blind height cmd.\\
Meteo parameters &        \\
Time                 &   \\
Closing time over 24h           &          \\
Seasonal Ext. temp.  &\\          
\bottomrule              
\end{tabular}
\end{sc}
\end{small}
\end{center}
\end{table}

The local room policies control the blinds and take values in the finite set $\{0,...,4\}$, the value $0$ being the value for opening the blind, and $4$ for fully closing it.  We applied the noise strategy for discrete action space described in Section \ref{optimal_training}.

The maps $\pi_g$ and $\pi_i$, $i \in \textrm{Rooms}$, are represented with neural networks. We used two types of neural networks: either a multilayer perceptron with two hidden layers, or an LSTM-network. The LSTM-networks can better keep track of the building/room states (see Section \ref{results}) but are more time intensive to train, in particular because time ordering has to be preserved in the mini-batches. We therefore chose to use only MLP networks for the blinds controllers and either MLP or LSTM networks for the global controller . As described in Section \ref{optimal_training}, we used a distributed approach for both local and global control, and trained the controllers with data collected by actors acting on different building models under distinct weather conditions. 

%Figure \ref{distributed_f} depicts an schematic  of the distributed framework.
%\begin{figure}[ht]
%\begin{center}
%\centerline{\includegraphics[scale=2]{EPFL_Buil_copies}}
%\caption{Distributed framework in the building test case. Reduced models are used in parallel at the slave nodes level.}
%\label{distributed_f}
%\end{center}
%\end{figure}
%\newline

%\noindent \textbf{Rewards.} 
\subsubsection*{Rewards}
Rewards were shaped differently for the local agent controllers and the global agent controller, following the soft constraint strategy explained in Section \ref{optimal_training}. For the local control of a room $i$, the reward takes the form
\begin{equation}
\begin{split}
r_i  &= 1 - \alpha_T  (T_i -23.0)^2  -\alpha_{c} t^{2}_{\textrm{closing}}\\
& - \alpha_{\max} \max (T_i-26,0)^{2} - \alpha_{\min} \max (19.5-T_i,0)^{2}, 
\end{split}
\end{equation}
where $\alpha_T, \alpha_{c},\alpha_{\max},\alpha_{\min} >0$ are used to tune the controller, $T_i$ is the room temperature, and $ t_{\textrm{closing}}$ is the closing time of the blind(s) over the last 24 hours. The penalization with $ t_{\textrm{closing}}$ was introduced to avoid convergence to a solution that would always keep the blinds closed, as it is the most efficient way to cool the building in summer time, but affects user comfort. 

For the global controller $\pi_g$, we chose a reward that takes into account the  mean temperature over all rooms, the mean temperature per floor and per ventilation zone and the minimum and maximum over all room temperatures. Moreover, a penalization was added for the energy consumption to optimize energy consumption under the temperature constraints. For the global controller, the reward takes the form
\begin{equation}
\begin{split}
r_g  &= 1 -  \sum_{f =0}^{2}\beta_{f}  (T_f -23.0)^2  -  \sum_{v=0}^{1}\beta_{v}  (T_v -23.0)^2\\
& - \beta_{\max}  \max_{i \in \textrm{Rooms}} (T_i-26,0)^{2} - \beta_{\min} \max_{i \in \textrm{Rooms}}  (19.5-T_i,0)^{2}\\
&  - \beta_E ( E_{\textrm{cooling}} + E_{\textrm{heating}}),
\end{split}
\end{equation}
where $T_f$ are the mean room floor temperatures, $T_v$ are the mean ventilation zone temperatures  and $ E_{\textrm{cooling}}$  and  $ E_{\textrm{heating}}$  are the thermal heating and cooling energies.
\newline 

%\noindent \textbf{Training strategy.} 
\subsubsection*{Training strategy}
Since there are 100 local controllers and one global controller, we chose first to train the local controllers with a global controller following the baseline. Then, the global controller was trained on top of the optimized local controllers, to improve global energy consumption and heating/cooling/ventilation distribution.

\subsection{Results summary}
\label{results}

%\noindent \textbf{Rule based controllers as baselines. }
\subsubsection*{Rule based controllers as baselines}
To benchmark the controller policies, we used  baseline controllers that  reproduce the historical control strategy of the building. 
Baseline controllers were fitted  with historical data and with the information provided by the building manager. To control the fluid temperature in the baseline model, we used a heating curve provided by the building manager. To control the change over (change between heating and cooling mode), we used the historical strategy consisting  in looking at the need for heating or cooling in the rooms: In each room, the difference between the measured  room temperature and the target temperature of 23 \degree C is evaluated. If the difference is higher than 0.5 \degree C or lower than -0.2 \degree C, the difference is added to the demand. Depending on the sign of the demand (positive or negative), the change over switches to heating or cooling mode. The ventilation supply temperature is used in addition to correct the mean room temperatures in the zone and to reach the ideal temperature of 23 \degree C. We used a linear model to fit the ventilation air supply temperature, with input variables the outside weather conditions, the change over position, and the temperature demands in the corresponding ventilation zone. Cross validation scores for the two zones gave an accuracy of 85\%. To fit the local valve control, we used the documentation provided by the supplier and a greedy policy search to identify the 2-way parameters of the valves from the nine months  of recorded historical data.
%\begin{figure}[ht]
%\begin{center}
%\centerline{\includegraphics[scale=0.4]{valve2537}}
%\caption{Example of fitted valve control for one of the room. A complete fit of the valves command is not possible only directly from the data, because the users can tune the desired temperature with a room thermostat: User changes affect the valves opening. }
%\label{r2}
%\end{center}
%\end{figure}
For the blinds baseline control, we used three strategies:   fixed schedules (four hours per day) based on room orientation, and smart schedules based on irradiance and room temperatures (see Appendix \ref{AppC} for more details).

%\noindent \textbf{Metrics.} 
\subsubsection*{Metrics}
We used three metrics to estimate the comfort felt by occupants in the building. First we estimated the absolute temperature difference in all rooms as compared to the ideal temperature of 23.0 \degree C: At  each timestep $t$, we computed
\begin{equation}
\textsc{dev}(t) := \sum_{r \in \textrm{rooms}} \vert T_{r}(t) -23.0 \vert.
\end{equation}
Next, we computed the maximum of $\textsc{dev}(t)$ over a day i.e.,  144 timesteps, and computed the mean of this value over all days in the available period $\mathbb{T}$:
\begin{equation}
\textsc{Mdev}= \frac{1}{ \# \mathbb{T}} \sum_{d \in \mathbb{T}} \max_{t \in[d,d+144] } \textsc{dev}(t),
\end{equation}
where $\# \mathbb{T}$ denotes the number of days in the period. Unfortunately, with the last metric, extreme temperatures that lead to  high discomfort over a short period of time are smeared out (the average is taken over all  rooms). Therefore, we used two other metrics, $N_{>26}$ and $N_{<20}$ that gives the total number of days spent in all rooms above 26 \degree C or below 20 \degree C.
\newline

%\noindent \textbf{Results.}
\subsubsection*{Results}
As real control was not possible due to building operator restrictions, we evaluated the trained controllers on simulations and compared them with the baselines over eight months of real meteorological data measured on the test site (February 2018 to October 2018). We used two reduced models of the building for the simulations: one closer to the historical data, obtained with full retraining, and, one closer to the simulation, obtained with partial retraining. None of these two reduced models had been experienced by the controllers during the training phase.

We named the controller models with the neural network used for the global control appended with the energy penalization coefficient. For example, the controller model $lstm_{en5}$  uses an LSTM for the global control and an energy coefficient penalization of 5. The higher this coefficient is, the more penalized the energy consumption is. Only MLP were used for the blind control, and the outputs of the blind controllers were regularized to avoid oscillatory behavior of the blinds i.e., blinds cannot change position more than two times over an hour. We use the abbreviation $C.E.$ and $H.E.$ for cooling and heating energies hereafter. 

\begin{figure}[t]
\begin{center}
\centerline{\includegraphics[scale=0.38]{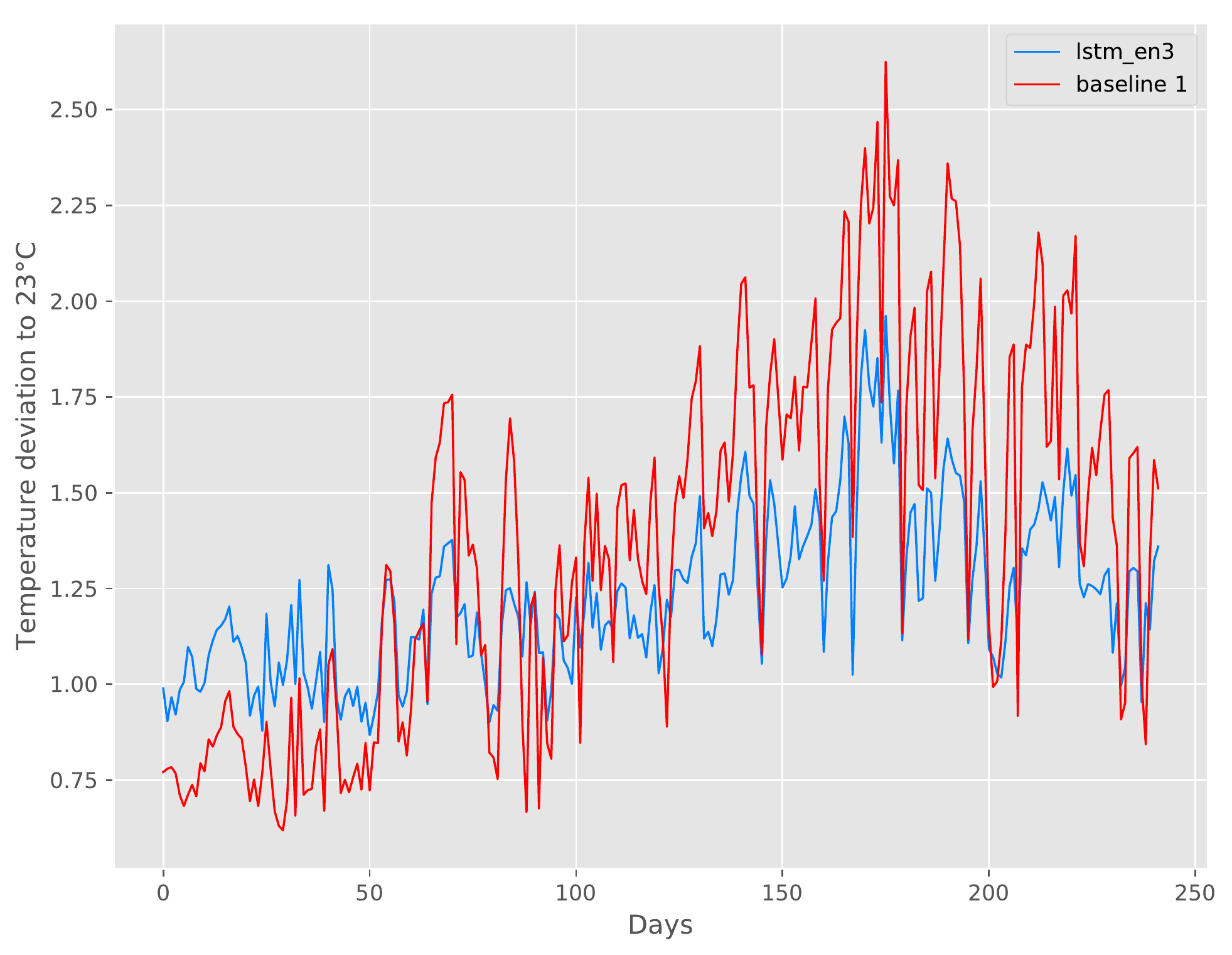}}
\caption{Example of comfort parameters $ \max_{t \in[d,d+144] } \textsc{dev}(t)$ for Baseline 1 and one trained controller. Evaluation during 240 days, on the reduced model building closer to the real data.}
\label{conf}
\end{center}
\end{figure}

The results obtained for the building closer to real data are presented in Table \ref{real_data_stats}. The best obtained results are marked in bold. No controller achieved the best scores in all the fields. However, LSTM-global controllers tend to score best. Figure \ref{conf} shows the evolution of the comfort parameter  $ \max_{t \in[d,d+144] } \textsc{dev}(t)$ for one of the learned controllers and Baseline 1 over 240 days. The maximum deviation of the baseline controller is lower than the learned controller during the first 50 days (winter) and then the learned controller achieves a lower deviation on average.
\begin{table}[h]
\caption{Results for building model closer to real data}
\label{real_data_stats}
\begin{center}
\begin{small}
\begin{sc}
\begin{tabular}{llllll}
\toprule
 \textbf{Controller}     &  Mdev  & C.E.& H.E.  & $N_{>26}$ & $N_{<20} $ \\
& {\tiny (\degree C)} & {\tiny  (kwh)}  & {\tiny  (kwh)}& {\tiny days} &  {\tiny days} \\
\midrule
$baseline \text{ 1}$&1.38&3.78&2.80&735&125\\
$baseline \text{ 2}$&1.53&3.83&2.80&932& \textbf{32}\\
$baseline \text{ 3}$&1.38 & 3.80 &2.80&519&171\\
$ mlp_{en0}$&1.20&4.50&2.36  &441&229       \\
$ mlp_{en3}$&1.26&4.78&1.83 &418&422   \\
$ mlp_{en5}$&1.26&4.87&2.02  &415&421         \\
$ mlp_{en9}$&1.29&4.01&1.88&568&257 \\
$ lstm_{en0}$&\textbf{1.16}&3.89&2.81 &440&60       \\
$ lstm_{en3}$&1.23&4.23&1.94  &446&256  \\
$ lstm_{en5}$&1.21&4.15&2.07   &445&224        \\
$ lstm_{en9}$&1.21&\textbf{3.20}&1.89 &\textbf{408}&360\\
$ lstm_{en11}$&1.28&4.37&\textbf{1.65}&436&358 \\ 
\bottomrule              
\end{tabular}
\end{sc}
\end{small}
\end{center}
\end{table}

We report in Table \ref{simu_data_stats} the results for the building closer to simulation. As for the other building model, the LSTM-controllers tend to outperform the MLP controllers. Surprisingly, energy consumption values for one particular controller are almost the same as for the model presented above. We found out that the commands were actually almost the same for the global controller but differed largely at the local level for the blinds. This may be due to two reasons:  the global controller is penalized by mean temperatures in the reward function, and, for the two building models, the critical observed inputs that correspond to the hottest zones (mean temperature on the second floor, mean temperature of the south zone) are actually very close. 
\begin{table}[ht]
\caption{Results for building model closer to simulation}
\label{simu_data_stats}
\begin{center}
\begin{small}
\begin{sc}
\begin{tabular}{llllll}
\toprule
 \textbf{Controller}     &  Mdev  & C.E.& H.E.  & $N_{>26}$ & $N_{<20} $ \\
& {\tiny (\degree C)} & {\tiny  (kwh)}  & {\tiny  (kwh)}& {\tiny days} &  {\tiny days} \\
\midrule
\midrule
$baseline \text{ 1}$&2.26&4.00&2.80&2231&24\\
$baseline \text{ 2}$&2.30&4.03&2.80&1874&\textbf{18}\\
$baseline \text{ 3}$&1.83&3.92&2.80&676&18\\
$ mlp_{en0}$& 1.48&4.49&2.36 &503&82 \\
$ mlp_{en3}$& 1.54&4.78&1.82&360&193   \\
$ mlp_{en5}$&1.53&4.87&2.01  &\textbf{358}&244         \\
$ mlp_{en9}$&1.74&4.01&1.88&747&55 \\
$ lstm_{en0}$& 1.49&3.87&2.82 &641&21      \\
$ lstm_{en3}$& 1.54&4.25&1.94 &514&44  \\
$ lstm_{en5}$&\textbf{1.47}&4.16&2.07&483&45          \\
$ lstm_{en9}$&1.91&\textbf{3.20}&1.89&1268&113 \\
$ lstm_{en11}$&1.62&4.37&\textbf{1.65}&453&137 \\ 
\bottomrule              
\end{tabular}
\end{sc}
\end{small}
\end{center}
\end{table}

\begin{figure}[th]
\begin{center}
\centerline{\includegraphics[scale=0.38]{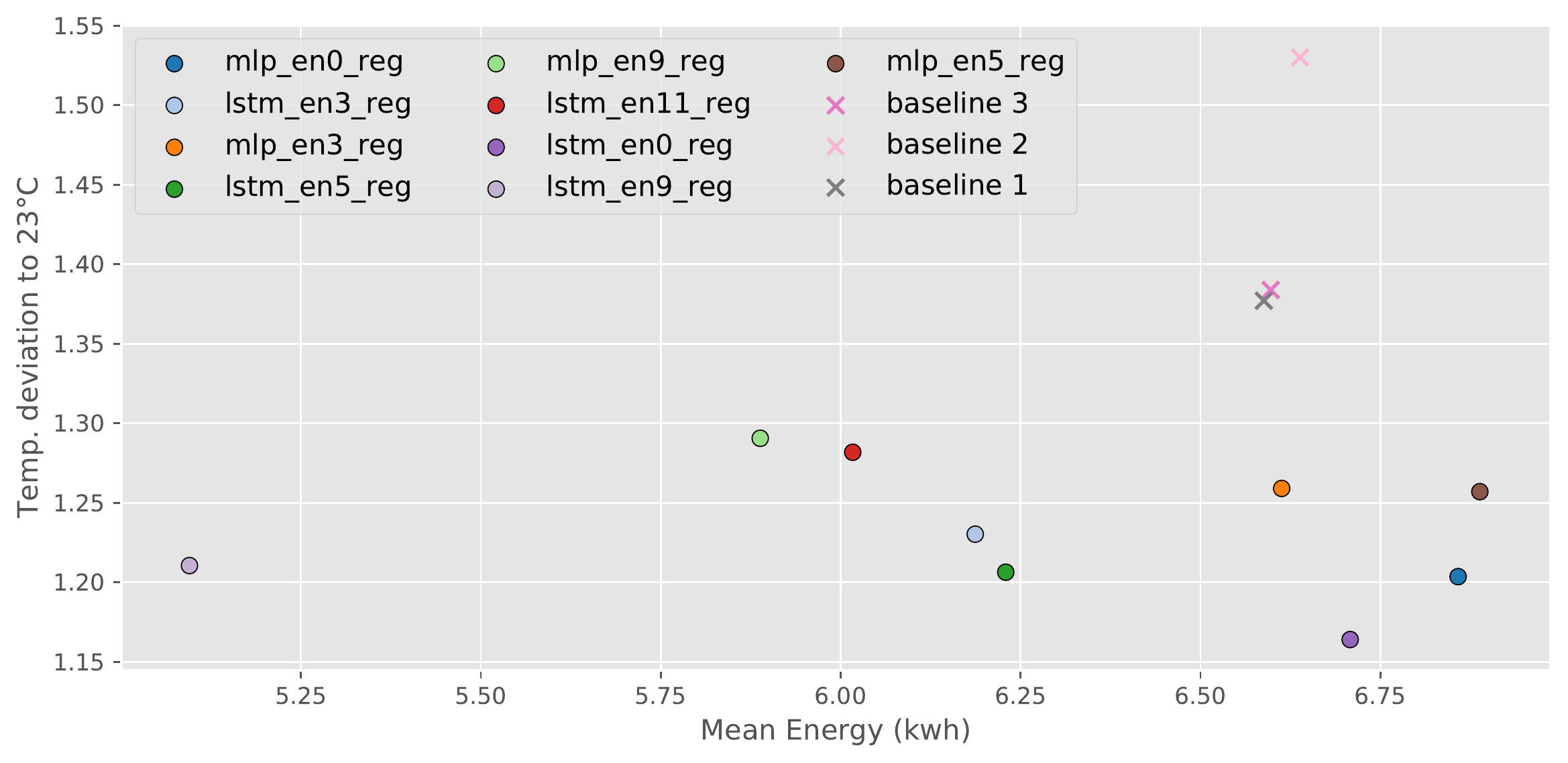}}
\caption{Total mean energy consumption vs. MDEV.}
\label{d_mean}
\end{center}
\end{figure}

%On Figure \ref{conf}  below, we plot again on a 2d-graph with the mean energy consumption (heating and cooling) versus the comfort index MDEV.  Notice that the LSTM model with energy coefficient 9 is that time performing less well on the comfort side, indicating that it may have seen during training more building models closer to real data than to simulation.
%\begin{figure}[ht]
%\begin{center}
%\centerline{\includegraphics[scale=0.38]{2d_en_con_700_01}}
%\caption{Total mean daily energy consumption vs. MDEV.}
%\label{d_mean}
%\end{center}
%\end{figure}
Figure \ref{d_mean} plots the mean energy consumption (heating and cooling) versus the comfort index MDEV. The trained controllers consistently outperform all baseline models for comfort and energy consumption over the year. A similar picture holds for the model closer to the simulation.

Blind control is key for reducing room temperatures because of the large glass facade of the building. As expected, longer daily closing duration leads to cooler room temperatures. It is difficult to find a good trade-off between closing blinds/ opening blinds and not disturbing the people working in the building. To reduce room temperatures, the learned controller tends to close the blinds during a longer time than the baseline. However, the closing  frequencies are quite similar. On average they both closed the blinds less than four hours a day: mean closing time per day was 3.26h for the proposed controller and 1.11h for baseline 3. This is still less than baseline 1, with a fixed four hours blind closing every day. However, in summer time, both controllers, and in particular the learned controller, tend to close the blinds longer. The median closing time was around six hours for the hottest days for the learned controller, meaning that half of the rooms had the blinds closed during more than 6 hours a day in the hottest period.
%
%\begin{figure}[ht]
%\begin{center}
%\centerline{\includegraphics[scale=0.38]{closing_time}}
%\centerline{\includegraphics[scale=0.38]{closing_freq}}
%\caption{Quantiles for the daily blinds closing time and positional change frequencies. }
%\label{clos-freq}
%\end{center}
%\end{figure}
The closing time of the optimizer may still be too long. It could be reduced by changing the penalization, with the drawback of degrading the room temperatures. On the other hand, we could control the blind opening angles by full closing to improve user comfort.

\section{Conclusion}
\label{conclusion}
We have presented a method to identify and control systems for which historical data are only available under closed-loop control and  system excitations  cannot be sent to the real system. We have shown that the system identification part of the method led to accurate and stable reduced models in both test cases, that can be unrolled with control commands and external parameters over a long time without accuracy degradation or divergences. When applied to the pendulum test case, the system identification method  outperforms all other implemented approaches. In the large building facility case, we were able to learn a reduced model of the entire building with median absolute error of 0.6 \degree C over all room temperatures, for three months ahead predictions. Furthermore, we have shown in simulations how a recent reinforcement learning approach can be used to learn a potentially optimal controller offline, with an ensemble of reduced models that are used to make the controller robust against parameters uncertainties  and noise. The proposed controllers outperform in terms of comfort and energy consumption the rule based controllers that we implemented as benchmarks and are close to the real control strategy implemented in the building. Unfortunately, we did not have the opportunity to evaluate our approach on the real building because of restrictions from the building operators (this illustrates, again, the difficulty of getting operators inclined to let optimal controllers work in real and complex buildings), though we plan to apply our method to other real systems in forthcoming work. We also emphasize that our technique could be useful for optimization of other energy systems.

One aspect that we did not address in the present paper is the following legitimate question:  Can we give theoretical bounds on the needed accuracy of the simulated dynamics, such that the system identification steps lead to a reduced model that is ``good enough'' for optimal control tasks? We plan to investigate  this question in a forthcoming work.

\section*{Acknowledgment}
We would like to thank Y. Stauffer, T. Gorecki and C. Jones for interesting discussions. We warmly thank P. Riederer for his help with getting in touch with  DIMOSIM. Part of this work was carried out in scope of the European Project THERMOSS. This work was supported by the Swiss State Secretariat for Education, Research and Innovation (SERI) under contract number 16.0106. The opinions expressed and arguments employed here do not necessarily reflect the official views of the Swiss Government.

% or

\appendices
\section{Hyperparameters}
\label{hyperparameters}
Hyperparameter values for the neural networks used in this article are listed in tables \ref{pendulum_hyper} and \ref{rooms_hyper}. Models were trained with Tensorflow 1.14 (see  \cite{tensorflow2015}).   For Algorithm \ref{partial}, the triangular shape was parametrized as in \cite{howar} with parameter values $cut\_frac = 0.1$, $\eta_{ \rm max}=0.1$, $ratio =40$, and $T= n_{\rm max }$.  
\begin{table}[ht]
\caption{Hyperparameter values for pendulum identification}
\label{pendulum_hyper}
\begin{center}
\begin{small}
\begin{sc}
\begin{tabular}{llllll}
\toprule
 Parameter  & Value\\
\midrule
\midrule
encoder length&12\\
decoder length&100\\
training batch&128\\
learning rate&1E-4 (adam)\\
LSTM cell size  ($p$ in Eq. \ref{rec} -\ref{rec2}) &32\\
MLP layer sizes & [64,32,16]\\
\bottomrule              
\end{tabular}
\end{sc}
\end{small}
\end{center}
\end{table}

\begin{table}[ht]
\caption{Hyperparameter values for rooms dynamics identification}
\label{rooms_hyper}
\begin{center}
\begin{small}
\begin{sc}
\begin{tabular}{llllll}
\toprule
 Parameter  & Value\\
\midrule
\midrule
encoder length&24\\
decoder length&144\\
training batch&200\\
learning rate&1E-4 (adam)\\
LSTM cell size  ($p$ in Eq. \ref{rec} -\ref{rec2})  & 256\\
MLP layer sizes & [256,128]\\
\bottomrule              
\end{tabular}
\end{sc}
\end{small}
\end{center}
\end{table}

For the local PPO controllers,  multilayer perceptrons with two internal layers of size 128 were used. For the global controller, either a multilayer perceptron with two layers of size 256 or an LSTM-network with cell size 128 were used.   Both types of controller were trained with Proximal Policy Optimization with the parameters displayed in Table \ref{PPO_hyper} (parameter names follow the original PPO article, \cite{schulman2017}). 

For the local controllers, training was carried out with 20 cores (i.e., 20 copies of the actors) in parallel.  Fixed-length trajectory segments of size $T =1440$ points where used at each iteration, i.e. the actors copies interacted 10 days with their attached reduced models (reduced models rollout time of 10 days) between two central weight updates. For the policy and critics update, minibatch sizes of $M=720$ points were sampled out  from the collected total batch rollout to compute stochastic gradient descent (SGD) of actor and critics at the central level.  Minibatch sampling and SGD were repeated for 2 epochs (i.e. L = 80 times) for each rollout. The process rollout  + SGD with minibatches was repeated 250 to 500 times.

For the global controller, training was also carried out with 20 cores (i.e., 20 copies of the actors) in parallel.  Fixed-length trajectory segments of size $T =576$ points where used at each iteration, i.e. the actors copies interacted 4 days with their attached reduced models between two central  weight updates. For the policy and critics update, minibatch sizes of $M=1152$ points were sampled out  from the collected total batch rollout to compute stochastic gradient descent  of actor and critics at the central level.  Minibatch sampling and SGD were repeated for 2 epochs (i.e. L = 80 times) for each rollout. The process rollout + SGD with minibatches was repeated 500 to 1000 times.

\begin{table}[H]
\caption{Hyperparameter values for PPO training}
\label{PPO_hyper}
\begin{center}
\begin{small}
\begin{sc}
\begin{tabular}{llllll}
\toprule
 Parameter  & Value\\
\midrule
\midrule
learning rate & 3E-4\\
clipping range&$0.1$\\
discount factor ($\gamma$)&0.99\\
GAE parameter ($\lambda$)&$0.95$\\
Value function coefficient ($c_1$)& $0.5$\\
Entropy coefficient ($c_2$) & $0.0$\\
\bottomrule              
\end{tabular}
\end{sc}
\end{small}
\end{center}
\end{table}

\section{Specifics of the building test case}
\label{AppC}
The building command set $\mathcal{C}$ for retraining was incorporating the following five days scenarios:
\begin{itemize}
\item Supply ventilation air at 20\degree C, blinds opened, valves for heating and cooling closed (S1).
\item Supply ventilation air at 20\degree C, blinds closed, valves for heating and cooling closed (S2).
\item Supply ventilation air at 20\degree C, blinds half closed, valves for heating and cooling opened with 20\degree C temperature fluid supply  (S3).
\item Supply ventilation air at 26 \degree C, blinds closed, valves opened and heating with temperature 50 \degree C (S4).
\item Supply ventilation air at 10 \degree C, blinds closed, valves opened and cooling with temperature 10 \degree C (S5).
\end{itemize}

Baselines used as benchmark had the following strategies for blinds control:
\begin{itemize}
\item Baseline 1:  Blinds are fully closed four hours a day at times depending on rooms orientation. Rooms with north/east orientation are closed between 8 am. to 12 am.. Rooms with south orientation are closed from 10 am. to 2 pm. and rooms with west orientation are closed at the end of the afternoon (2 pm. to 6 pm).
\item Baseline 2: Blinds are fully closed if the room temperature is above 25.5 \degree C and the outside irradiance is larger than $400\text{ } W.m^{-2}$.  Blinds are half closed if the room temperature is above 24 \degree C  and the measured irradiance is larger than $400\text{ } Wm^{-2}$.
\item Baseline 3:  Blinds are fully closed if the room temperature is above 24.5 \degree C and the outside irradiance is larger than $300\text{ } W.m^{-2}$.  Blinds are half closed if the room temperature is above 24 \degree C  and the measured irradiance is larger than $400\text{ } Wm^{-2}$.
\end{itemize}

\bibliography{thermoss_tnnls_final}

\begin{thebibliography}{10}
\providecommand{\url}[1]{#1}
\csname url@samestyle\endcsname
\providecommand{\newblock}{\relax}
\providecommand{\bibinfo}[2]{#2}
\providecommand{\BIBentrySTDinterwordspacing}{\spaceskip=0pt\relax}
\providecommand{\BIBentryALTinterwordstretchfactor}{4}
\providecommand{\BIBentryALTinterwordspacing}{\spaceskip=\fontdimen2\font plus
\BIBentryALTinterwordstretchfactor\fontdimen3\font minus
  \fontdimen4\font\relax}
\providecommand{\BIBforeignlanguage}[2]{{%
\expandafter\ifx\csname l@#1\endcsname\relax
\typeout{** WARNING: IEEEtran.bst: No hyphenation pattern has been}%
\typeout{** loaded for the language `#1'. Using the pattern for}%
\typeout{** the default language instead.}%
\else
\language=\csname l@#1\endcsname
\fi
#2}}
\providecommand{\BIBdecl}{\relax}
\BIBdecl

\bibitem{camacho2007}
E.~F. Camacho and C.~Bordons, \emph{Model predictive control}.\hskip 1em plus
  0.5em minus 0.4em\relax Springer, 2007.

\bibitem{di2012industry}
S.~Di~Cairano, ``An industry perspective on mpc in large volumes applications:
  Potential benefits and open challenges,'' \emph{IFAC Proceedings Volumes},
  vol.~45, no.~17, pp. 52--59, 2012.

\bibitem{van2012subspace}
P.~Van~Overschee and B.~De~Moor, \emph{Subspace identification for linear
  systems: Theory—Implementation—Applications}.\hskip 1em plus 0.5em minus
  0.4em\relax Springer Science \& Business Media, 2012.

\bibitem{borrelli2017predictive}
F.~Borrelli, A.~Bemporad, and M.~Morari, \emph{Predictive control for linear
  and hybrid systems}.\hskip 1em plus 0.5em minus 0.4em\relax Cambridge
  University Press, 2017.

\bibitem{mayne2014}
D.~Q. Mayne, ``Model predictive control: Recent developments and future
  promise,'' \emph{Automatica}, vol.~50, no.~12, pp. 2967 -- 2986, 2014.

\bibitem{schoukens2019nonlinear}
J.~Schoukens and L.~Ljung, ``Nonlinear system identification: A user-oriented
  roadmap,'' \emph{arXiv preprint arXiv:1902.00683}, 2019.

\bibitem{jones2007air}
W.~Jones, \emph{Air conditioning engineering}.\hskip 1em plus 0.5em minus
  0.4em\relax Routledge, 2007.

\bibitem{schulman2017}
J.~Schulman, F.~Wolski, P.~Dhariwal, A.~Radford, and O.~Klimov, ``Proximal
  policy optimization algorithms,'' \emph{arXiv preprint arXiv:1707.06347},
  2017.

\bibitem{ppo_18}
M.~Andrychowicz, B.~Baker, M.~Chociej, R.~Jozefowicz, B.~McGrew, J.~Pachocki,
  A.~Petron, M.~Plappert, G.~Powell, A.~Ray, and all, ``Learning dexterous
  in-hand manipulation,'' \emph{arXiv preprint arXiv:1808.00177}, 2018.

\bibitem{tobin2017domain}
J.~Tobin, R.~Fong, A.~Ray, J.~Schneider, W.~Zaremba, and P.~Abbeel, ``Domain
  randomization for transferring deep neural networks from simulation to the
  real world,'' in \emph{2017 IEEE IROS}.\hskip 1em plus 0.5em minus
  0.4em\relax IEEE, 2017, pp. 23--30.

\bibitem{peng2018sim}
X.~Peng, M.~Andrychowicz, W.~Zaremba, and P.~Abbeel, ``Sim-to-real transfer of
  robotic control with dynamics randomization,'' in \emph{2018 IEEE
  ICRA}.\hskip 1em plus 0.5em minus 0.4em\relax IEEE, 2018, pp. 1--8.

\bibitem{dai2017sbeed}
B.~Dai, A.~Shaw, L.~Li, L.~Xiao, N.~He, Z.~Liu, J.~Chen, and L.~Song, ``Sbeed:
  Convergent reinforcement learning with nonlinear function approximation,''
  \emph{arXiv preprint arXiv:1712.10285}, 2017.

\bibitem{liu2019neural}
B.~Liu, Q.~Cai, Z.~Yang, and Z.~Wang, ``Neural proximal/trust region policy
  optimization attains globally optimal policy,'' \emph{arXiv preprint
  arXiv:1906.10306}, 2019.

\bibitem{ljung2001system}
L.~Ljung, ``System identification,'' \emph{Wiley Encyclopedia of Electrical and
  Electronics Engineering}, 2001.

\bibitem{schoukens2016}
J.~{Schoukens}, M.~{Vaes}, and R.~{Pintelon}, ``Linear system identification in
  a nonlinear setting: Nonparametric analysis of the nonlinear distortions and
  their impact on the best linear approximation,'' \emph{IEEE Control Systems
  Magazine}, vol.~36, no.~3, pp. 38--69, June 2016.

\bibitem{nechyba1994}
M.~C. {Nechyba} and {Yangsheng Xu}, ``Neural network approach to control system
  identification with variable activation functions,'' in \emph{Proceedings of
  1994 9th IEEE International Symposium on Intelligent Control}, Aug 1994, pp.
  358--363.

\bibitem{ogunmolu2016}
O.~P. Ogunmolu, X.~Gu, S.~B. Jiang, and N.~R. Gans, ``Nonlinear systems
  identification using deep dynamic neural networks,'' \emph{ArXiv}, vol.
  abs/1610.01439, 2016.

\bibitem{gonzales2018}
J.~Gonzalez and W.~Yu, ``Non-linear system modeling using lstm neural
  networks,'' \emph{IFAC-PapersOnLine}, vol.~51, no.~13, pp. 485 -- 489, 2018,
  2nd IFAC Conference on Modelling, Identification and Control of Nonlinear
  Systems MICNON 2018.

\bibitem{baumeister2018deep}
T.~Baumeister, S.~L. Brunton, and J.~Kutz, ``Deep learning and model predictive
  control for self-tuning mode-locked lasers,'' \emph{JOSA B}, vol.~35, no.~3,
  pp. 617--626, 2018.

\bibitem{champion2019data}
K.~Champion, B.~Lusch, J.~Kutz, and S.~Brunton, ``Data-driven discovery of
  coordinates and governing equations,'' \emph{PNAS}, vol. 116, no.~45, pp.
  22\,445--22\,451, 2019.

\bibitem{lenz2015}
I.~Lenz, R.~A. Knepper, and A.~Saxena, ``Deepmpc: Learning deep latent features
  for model predictive control,'' in \emph{Robotics: Science and Systems},
  2015.

\bibitem{drgona2018}
J.~Drgoňa, D.~Picard, M.~Kvasnica, and L.~Helsen, ``Approximate model
  predictive building control via machine learning,'' \emph{Applied Energy},
  vol. 218, no.~23, pp. 199--216, 2018.

\bibitem{verma2018}
S.~Verma, G.~Novati, and P.~Koumoutsakos, ``Efficient collective swimming by
  harnessing vortices through deep reinforcement learning,'' \emph{Proceedings
  of the National Academy of Sciences}, vol. 115, no.~23, pp. 5849--5854, 2018.

\bibitem{bieker2019}
K.~Bieker, S.~Peitz, S.~L. Brunton, J.~N. Kutz, and M.~Dellnitz, ``Deep model
  predictive control with online learning for complex physical systems,''
  \emph{ArXiv}, vol. abs/1905.10094, 2019.

\bibitem{serale2018}
G.~Serale, M.~Fiorentini, A.~Capozzoli, D.~Bernardini, and A.~Bemporad, ``Model
  predictive control (mpc) for enhancing building and hvac system energy
  efficiency: Problem formulation, applications and opportunities,''
  \emph{Energies}, vol.~11, no.~3, 2018.

\bibitem{mocanu2019}
E.~{Mocanu}, D.~C. {Mocanu}, P.~H. {Nguyen}, A.~{Liotta}, M.~E. {Webber},
  M.~{Gibescu}, and J.~G. {Slootweg}, ``On-line building energy optimization
  using deep reinforcement learning,'' \emph{IEEE Transactions on Smart Grid},
  vol.~10, no.~4, pp. 3698--3708, July 2019.

\bibitem{lindelof2015}
D.~Lindelof, H.~Afshari, M.~Alisafaee, J.~Biswas, M.~Caban, X.~Mocellin, and
  J.~Viaene, ``Field tests of an adaptive,model-predictive heating controller
  for residential buildings,'' \emph{Energy and Buildings}, vol.~99, pp.
  292--302, 2015.

\bibitem{sutton2018reinforcement}
R.~S. Sutton and A.~Barto, \emph{Reinforcement learning: An
  introduction}.\hskip 1em plus 0.5em minus 0.4em\relax MIT press, 2018.

\bibitem{mnih2015human}
V.~Mnih \emph{et~al.}, ``Human-level control through deep reinforcement
  learning,'' \emph{Nature}, vol. 518, no. 7540, p. 529, 2015.

\bibitem{silver2016mastering}
D.~Silver \emph{et~al.}, ``Mastering the game of go with deep neural networks
  and tree search,'' \emph{nature}, vol. 529, no. 7587, p. 484, 2016.

\bibitem{vinyals2019alphastar}
O.~Vinyals \emph{et~al.}, ``Alphastar: Mastering the real-time strategy game
  starcraft ii,'' \emph{DeepMind Blog}, 2019.

\bibitem{schaal1999imitation}
S.~Schaal, ``Is imitation learning the route to humanoid robots?'' \emph{Trends
  in cognitive sciences}, vol.~3, no.~6, pp. 233--242, 1999.

\bibitem{osa2018algorithmic}
T.~Osa \emph{et~al.}, ``An algorithmic perspective on imitation learning,''
  \emph{Foundations and Trends in Robotics}, vol.~7, no. 1-2, pp. 1--179, 2018.

\bibitem{vezzani2019learning}
G.~Vezzani, A.~Gupta, L.~Natale, and P.~Abbeel, ``Learning latent state
  representation for speeding up exploration,'' \emph{arXiv preprint
  arXiv:1905.12621}, 2019.

\bibitem{janner2019trust}
M.~Janner, J.~Fu, M.~Zhang, and S.~Levine, ``When to trust your model:
  Model-based policy optimization,'' \emph{arXiv preprint arXiv:1906.08253},
  2019.

\bibitem{bueno2012resistance}
B.~Bueno, L.~Norford, G.~Pigeon, and R.~Britter, ``A resistance-capacitance
  network model for the analysis of the interactions between the energy
  performance of buildings and the urban climate,'' \emph{Building and
  Environment}, vol.~54, pp. 116--125, 2012.

\bibitem{lstm_art}
S.~Hochreiter and J.~Schmidhuber, ``Long short-term memory,'' \emph{Neural
  computation}, vol.~9, no.~8, pp. 1735--1780, 1997.

\bibitem{NIPS2014_5346}
I.~Sutskever, O.~Vinyals, and Q.~V. Le, ``Sequence to sequence learning with
  neural networks,'' in \emph{Advances in Neural Information Processing Systems
  27}.\hskip 1em plus 0.5em minus 0.4em\relax Curran Associates, Inc., 2014,
  pp. 3104--3112.

\bibitem{cho2014properties}
K.~Cho, B.~Van~Merri{\"e}nboer, D.~Bahdanau, and Y.~Bengio, ``On the properties
  of neural machine translation: Encoder-decoder approaches,'' \emph{arXiv
  preprint arXiv:1409.1259}, 2014.

\bibitem{vaswani2017attention}
A.~Vaswani, N.~Shazeer, N.~Parmar, J.~Uszkoreit, L.~Jones, A.~Gomez, L.~Kaiser,
  and I.~Polosukhin, ``Attention is all you need,'' in \emph{Advances in neural
  information processing systems}, 2017, pp. 5998--6008.

\bibitem{fateh2019state}
A.~Fateh, D.~Borelli, A.~Spoladore, and F.~Devia, ``A state-space analysis of a
  single zone building considering solar radiation, internal radiation, and pcm
  effects,'' \emph{Applied Sciences}, vol.~9, no.~5, p. 832, 2019.

\bibitem{zhang2019gradient}
J.~Zhang, T.~He, S.~Sra, and A.~Jadbabaie, ``Why gradient clipping accelerates
  training: A theoretical justification for adaptivity,'' in \emph{ICLR}, 2019.

\bibitem{peters2013}
J.~Peters, D.~Janzing, and B.~Sch{\"o}lkopf, ``Causal inference on time series
  using restricted structural equation models,'' in \emph{NIPS}, 2013, pp.
  154--162.

\bibitem{howar}
J.~Howard and S.~Ruder, ``Universal language model fine-tuning for text
  classification,'' in \emph{Proceedings of the 56th Annual Meeting of the
  Association for Computational Linguistics (Volume 1: Long Papers)}.\hskip 1em
  plus 0.5em minus 0.4em\relax Melbourne, Australia: Association for
  Computational Linguistics, Jul. 2018, pp. 328--339.

\bibitem{schulman2015trust}
J.~Schulman, S.~Levine, P.~Abbeel, M.~Jordan, and P.~Moritz, ``Trust region
  policy optimization,'' in \emph{ICML}, 2015, pp. 1889--1897.

\bibitem{heess2017}
N.~Heess, S.~Sriram, J.~Lemmon, J.~Merel, G.~Wayne, Y.~Tassa, T.~Erez, Z.~Wang,
  A.~Eslami, M.~Riedmiller \emph{et~al.}, ``Emergence of locomotion behaviours
  in rich environments,'' \emph{arXiv preprint arXiv:1707.02286}, 2017.

\bibitem{perez2015thermal}
N.~Perez, P.~Riederer, C.~Inard, and V.~Partenay, ``Thermal building modeling
  adapted to district energy simulation,'' \emph{Proceedings of BS2015}, 2015.

\bibitem{CSTB}
\BIBentryALTinterwordspacing
 [Online]. Available: \url{http://www.cstb.fr/fr/}
\BIBentrySTDinterwordspacing

\bibitem{tensorflow2015}
\BIBentryALTinterwordspacing
M.~Abadi \emph{et~al.}, ``{TensorFlow}: Large-scale machine learning on
  heterogeneous systems,'' 2015, software available from tensorflow.org.
  [Online]. Available: \url{https://www.tensorflow.org/}
\BIBentrySTDinterwordspacing

\end{thebibliography}
\bibliographystyle{IEEEtran}


%\end{IEEEbiography}


%\newpage



%\vfill


\end{document}